\documentclass[sigconf]{acmart}

\AtBeginDocument{%
  \providecommand\BibTeX{{%
    \normalfont B\kern-0.5em{\scshape i\kern-0.25em b}\kern-0.8em\TeX}}}

\copyrightyear{2025}
\acmYear{2025}
\setcopyright{acmlicensed}
\acmConference[KDD '25] {Proceedings of the 31st ACM SIGKDD Conference on Knowledge Discovery and Data Mining V.2}{August 3--7, 2025}{Toronto, ON, Canada}
\acmBooktitle{Proceedings of the 31st ACM SIGKDD Conference on Knowledge Discovery and Data Mining V.2 (KDD '25), August 3--7, 2025, Toronto, ON, Canada}
\acmISBN{979-8-4007-1454-2/2025/08}
\acmDOI{10.1145/3711896.3737026}

\usepackage[T1]{fontenc}
\usepackage{aecompl}
\usepackage{booktabs}
\usepackage{amsmath,amsfonts}

\usepackage{algorithmic}
\usepackage{graphicx}
\usepackage{subfigure}
\usepackage{textcomp}
\usepackage{framed}
\usepackage{bm}
\usepackage{enumerate}
\usepackage{tcolorbox}
\usepackage{xcolor}
\usepackage{arydshln}
\usepackage{multirow}
\usepackage{hyperref}
\usepackage[export]{adjustbox}
\usepackage{enumitem}
\usepackage{microtype}
\usepackage{balance}
\usepackage{makecell}
\usepackage{todonotes}

\usepackage{subfigure}
\usepackage{pgfplots}
\makeatletter
\renewcommand{\@thesubfigure}{\hskip\subfiglabelskip}
\makeatother

\setitemize[1]{itemsep=0pt,partopsep=0pt,parsep=\parskip,topsep=5pt}

\usepackage[ruled,linesnumbered]{algorithm2e}

\begin{document}
\title{LightKG: Efficient Knowledge-Aware Recommendations with Simplified GNN Architecture}

\author{Yanhui Li}
\orcid{0009-0000-2180-9157}
\affiliation{%
  \institution{Zhejiang University}
  \city{Hangzhou}
  \country{China}
}
\email{YanHuiLi@zju.edu.cn}

\author{Dongxia Wang}
\orcid{0000-0001-9812-3911}
\authornote{Corresponding author.}
\affiliation{%
  \institution{Zhejiang University}
  \city{Hangzhou}
  \country{China}
}
\email{dxwang@zju.edu.cn}

\author{Zhu Sun}
\orcid{0000-0002-3350-7022}
\affiliation{%
  \institution{Singapore University of Technology and Design}
  \country{Singapore}
}
\email{sunzhuntu@gmail.com}

\author{Haonan Zhang}
\orcid{0009-0009-9782-4121}
\affiliation{%
  \institution{Zhejiang University}
  \city{Hangzhou}
  \country{China}
}
\email{haonanzhang@zju.edu.cn}

\author{Huizhong Guo}
\orcid{0009-0004-0011-8612}
\affiliation{%
  \institution{Zhejiang University}
  \city{Hangzhou}
  \country{China}
}
\email{huiz_g@zju.edu.cn}

\renewcommand{\shortauthors}{YanHui Li, Dongxia Wang, Zhu Sun, Haonan Zhang, and Huizhong Guo}


\begin{abstract} 
Recently, Graph Neural Networks (GNNs) have become the dominant approach for Knowledge Graph-aware Recommender Systems (KGRSs) due to their proven effectiveness. Building upon GNN-based KGRSs, Self-Supervised Learning (SSL) has been incorporated to address the sparity issue, leading to longer training time. However, through extensive experiments, we reveal that: (1)compared to other KGRSs, the existing GNN-based KGRSs fail to keep their superior performance under sparse interactions even with SSL. (2) More complex models tend to perform worse in sparse interaction scenarios and  complex mechanisms, like attention mechanism, can be detrimental as they often increase learning difficulty. Inspired by these findings, we propose LightKG, a simple yet powerful GNN-based KGRS to address sparsity issues. LightKG includes a simplified GNN layer that encodes directed relations as scalar pairs rather than dense embeddings and employs a linear aggregation framework, greatly reducing the complexity of GNNs. Additionally, LightKG incorporates an efficient contrastive layer to implement SSL. It directly minimizes the node similarity in original graph, avoiding the time-consuming subgraph generation and comparison required in previous SSL methods. Experiments on four benchmark datasets show that LightKG outperforms 12 competitive KGRSs in both sparse and dense scenarios while significantly reducing training time. Specifically, it surpasses the best baselines by an average of 5.8\% in recommendation accuracy and saves 84.3\% of training time compared to KGRSs with SSL. Our code is available at \url{https://github.com/1371149/LightKG}.
\end{abstract}



\maketitle

\section{Introduction}
Recommender Systems (RSs) aim to capture user preferences from their behavior to filter irrelevant information in domains such as movies, news, and e-commerce~\cite{jiang2024diffkg}.
Due to the Matthew effect~\cite{chen2024opportunities}, interactions in real-world scenarios often concentrate on a few popular items or active users, resulting in "sparse scenarios" where the majority of users and items have limited interactions.
Such scenarios pose significant challenges to RSs in providing accurate and personalized recommendations, where poor performance could lead to user churn and significant economic losses~\cite{shi2024enhancing}. 

To address this, Knowledge Graphs (KGs) have been incorporated into RSs to provide rich auxiliary information, forming Knowledge Graph-aware Recommender Systems (KGRSs)~\cite{wen2024multi}.
Based on the model framework, KGRSs can be categorized into three types, among which Graph Neural Network (GNN) emerges as the dominant solution~\cite{li2023survey}, primarily due to their proven effectiveness in encoding hierarchical relational patterns in KGs.
Recent advancements leverage Self-Supervised Learning (SSL)~\cite{zou2022multi,yang2022knowledge} to extract more supervisory signals, enhancing recommendation performance in sparse scenarios while leading to longer training time (see Tab.~\ref{Efficiency_Comparison}).

\begin{table}[tp]
\setlength{\abovecaptionskip}{3pt}
\centering
\caption{
The Recall@10 and training time per epoch for GNN-based KGRSs. More recently-proposed models tend to have longer training time, indicating increased model complexity.}
\vspace{-0.01in}
\label{Efficiency_Comparison}
\resizebox{0.975\linewidth}{!}
{
\begin{tabular}{cccccccccc}
\toprule
\multicolumn{1}{l}{} & \multirow{2}{*}{Model} & \multicolumn{2}{c}{Amazon-book} & \multicolumn{2}{c}{ML-1M} & \multicolumn{2}{c}{Book-Crossing} & \multicolumn{2}{c}{Last.FM} \\ 
\multicolumn{1}{l}{} &  & Recall & time & Recall & time & Recall & time & Recall & time \\ \midrule
\multirow{4}{1.2cm}{Supervised method} & KGCN(2019) & 0.1550 & 0.76s & 0.1594 & 2.13s & 0.0467 & 0.72s & 0.2149 & 0.71s \\
 & KGNN-LS(2019) & 0.1508 & 1.52s & 0.1592 & 5.92s & 0.0330 & 1.50s & 0.2117 & 1.33s \\
 & KGAT(2019) & 0.1925 & 1.78s & 0.1830 & 8.73s & 0.0499 & 1.81s & 0.2583 & 0.80s \\
 & KGIN(2021) & 0.2090 & 8.21s & 0.1969 & 19.26s & 0.0801 & 10.21s & 0.2727 & 1.48s \\ \midrule
\multirow{5}{1.2cm}{Self-Supervised method} & MCCLK(2022) & 0.2025 & 8.48s & 0.1853 & 161.48s & 0.0607 & 59.06s & 0.2699 & 5.01s \\
 & KGRec(2023) & 0.2035 & 4.47s & 0.1960 & 99.69s & 0.1033 & 21.15s & 0.2560 & 3.10s \\
 & DiffKG(2024) & 0.2039 & 16.87s & 0.1846 & 115.46s & 0.0581 & 60.92s & 0.2520 & 2.58s \\
 & CL-SDKG(2024) & 0.2036 & 52.52s & 0.1861 & 281.2s & 0.0924 & 9.39s & 0.2409 & 1.92S \\ 
 & LightKG(Ours) & 0.2120 & 3.70s & 0.2029 & 19.07s & 0.1154 & 1.58s & 0.2929 & 0.92s \\
 \bottomrule
\end{tabular}
}
\vspace{-0.6cm}
\end{table}

However, we evaluate the recommendation accuracy of 12 KGRSs under varying levels of interaction sparsity and observe that \emph{while 
GNN-based KGRSs always achieve the best performance compared to other KGRSs in dense scenarios, they generally fail to do so in sparse scenario even with SSL.}
Then, we analyze the complexity of each GNN-based KGRS and find that \emph{more complex models tend to perform worse in sparse scenarios}.
To further explore the influence of model complexity, we then design experiments to investigate whether and how simplifying GNN-based KGRSs, e.g., removing the attention mechanism employed by six representative KGRSs (e.g., KGAT~\cite{wang2019kgat}), would affect their recommendation accuracy. 
Surprisingly, the results show that \emph{the removal mostly influences little, while often even increases the accuracy slightly}.  
These observations suggest that complex mechanisms in GNN-based KGRSs may be unsuitable for sparse scenarios, potentially leading to overfitting or poor generalization.
This insight inspires us to look for a simpler yet effective GNN-based KGRS to address the sparsity issue.

To this end, we propose LightKG, a simple yet powerful GNN-based KGRS, to  address the sparsity issue.
It includes a \emph{simplified GNN layer} to derive node embeddings and an \emph{efficient contrastive layer} to implement SSL from the perspective of alleviating the over-smoothing issue.
In the GNN layer, unlike other KGRSs that encode relations as vectors or transformation matrices, LightKG encodes them simply as scalar pairs. 
Then, LightKG employs these scalars to a linear aggregation framework, which significantly reduces the model's complexity. 
In the contrastive layer, LightKG avoids the generation and comparison of subgraphs, which is time-consuming required in previous SSL methods. Instead, LightKG directly minimizes the similarity between nodes on the original graph, which significantly reduces the training time for SSL.

Equipped with a simpler scalar-based relation encoding strategy, a linear
information aggregation framework, and efficient contrastive layers, we examine how LightKG performs compared to the existing KGRSs. 
We apply LightKG to four benchmark datasets which vary greatly in sizes and sparsity levels.
The experimental results show that LightKG outperforms 12 State-Of-The-Art (SOTA) KGRSs of diverse types on these benchmark datasets, not only with sparse but also with dense interactions.
Meanwhile, LightKG shows high training efficiency compared to other KGRSs with SSL.
To summarize, our contributions are highlighted as follows: 
\vspace{-0.1cm}
\begin{itemize}[leftmargin=0.3cm]
\item For the first time, we empirically reveal that simpler GNN-based KGRSs can perform better with sparse interactions, whereas complex mechanisms like attention in GNNs have minimal impact or may even degrade the recommendation accuracy.

\item  We propose LightKG, a simple yet powerful GNN-based KGRS which incorporates a much simpler relation encoding strategy, a linear information aggregation framework and efficient contrastive layers. These designs allow LightKG to significantly reduce training time while enhancing recommendation accuracy, particularly in sparse scenarios.

\item We evaluate LightKG on four benchmark datasets against 12 SOTA KGRSs. 
The results demonstrate its superiority in both recommendation accuracy and training efficiency. Particularly, it surpasses the best baseline by margins ranging from $1.4\%$ to $11.7\%$ on accuracy, and requires only $16.8\%$ to $82.7\%$ of the training time compared to SOTA GNN-based KGRSs with SSL.
\end{itemize}

\section{Preliminary}
In this section, we will briefly introduce the related concepts, including how a GNN-based KGRS generally works and the Collaborative Knowledge Graph (CKG)~\cite{wang2019kgat} it usually employs.

Usually a GNN-based KGRS is trained on a CKG, which consists of a user-item interaction graph $\mathcal{G}_{UI}$ and a KG $\mathcal{G}_{KG}$.
For example, as shown in Fig.~\ref{fig:framework}, \textit{Tom buys a book written by Jane}, where \textit{Tom} is a user, the \textit{book} is an item, the author \textit{Jane} is an entity, and the relation $r$ between \textit{Jane} and the \textit{book} is ‘book-author’.
Specifically, let $\mathcal{U},\mathcal{I},\mathcal{V}$  denote the set of users, items and entities respectively.
$\mathcal{G}_{UI} = \{ (u,\textit{Interact},i)|u \in  \mathcal{U},i \in \mathcal{I}\}$, where \textit{Interact} is a relation, meaning an observed interaction between user $u$ and item $i$. 
$\mathcal{G}_{KG}=\{ (h,r,t)| h,t \in (\mathcal{I} \cup \mathcal{V}), r \in \mathcal{R'}  \}$, where  $\mathcal{R}'$ denotes the set of relations in KG ($\textit{Interact} \notin \mathcal{R}'$).
The user-item graph can be integrated with a KG as a CKG: $\mathcal{G}=\{(h,r,t)|h,t \in \mathcal{K}, r \in \mathcal{R}\}$, where $\mathcal{K} =  (\mathcal{U} \cup \mathcal{I} \cup \mathcal{V})$, $\mathcal{R}=\mathcal{R'} \cup \{ \textit{Interact}\}$.

To train a model, each node and relation will firstly be mapped to the feature space.
Let $\boldsymbol{e}^{(0)}_{k}$ denote the embedding of node $k$, and $\boldsymbol{e}_r$ denote the embedding of relation $r$.
Let $\mathcal{N}_{k}$ denote the set of neighbors of node $k$.
The node embeddings are randomly initialized. 
Then with the involvement of the relations, the model updates the node embeddings by aggregating information from its neighboring nodes, performing it iterately for $L$ times as shown in Equ.~\ref{eq:agg}. 
Finally, a more accurate node embedding is obtained for RSs, given by, 
\begin{equation}
\footnotesize
\label{eq:agg}
\boldsymbol{e}_{k}^{(l)} = \sum\nolimits_{(k,r,t) \in \mathcal{N}_{k}} \boldsymbol{Agg}(\boldsymbol{e}_{t}^{(l-1)},\boldsymbol{e}_r), l \in {1,2,...,L},
\end{equation}
where $\boldsymbol{Agg(\cdot)}$ is a function used for embedding aggregation, which varies in different models.
Embedding $\boldsymbol{e}_r$ can be encoded as vectors or matrices depending on the specific GNN-based KGRS.
Together with the embeddings obtained by the aggregation in the $L$ layers, we obtain $L+1$ embeddings for node $k$: $\{ \boldsymbol{e}_{k}^{(0)},\boldsymbol{e}_{k}^{(1)},\boldsymbol{e}_{k}^{(2)},...,\boldsymbol{e}_{k}^{(L)}\}$.
The embedding of node $k$ for model prediction, i.e., $\boldsymbol{e}_{k}^{*}$ can be obtained using a combination function,
\begin{equation}
\footnotesize
\label{eq:comb}
    \boldsymbol{e}_{k}^{*} = \boldsymbol{Comb}(\boldsymbol{e}_{k}^{(0)},\boldsymbol{e}_{k}^{(1)},\boldsymbol{e}_{k}^{(2)},...,\boldsymbol{e}_{k}^{(L)}).
\end{equation}
Finally, the embeddings $\boldsymbol{e}_u^{*}$, $\boldsymbol{e}_{i}^{*}$ of each user-item pair $(u,i)$ are used to get the prediction score $\hat{y}_{ui} = \boldsymbol{e}_{u}^{*\top} \boldsymbol{e}_{i}^{*}$. 
Based on these prediction scores, all candidate items are ranked in descending order. The top-ranked items are then selected and recommended to the users. 

\input{photo/sparse_experiment}

\section{Motivation}\label{sec:motivation}
In this section, we present a series of exploratory experiments, the observations of which motivate our proposal for LightKG.
We focus on scenarios where user-item interaction is sparse.
First, we evaluate 12 KGRSs at different sparsity levels and observed that \emph{while GNN-based KGRSs always achieve the best performance compared to other KGRSs in dense scenarios, they fail to do so in sparse scenario}.
Then, we analyze the complexity of each GNN model and \emph{find that higher-complexity models tend to perform poorly in sparse scenarios}.
To further explore the connection between complexity and performance under sparse scenarios, we simplify several SOTA KGRSs by removing their attention mechanisms. 
Such removal yields a surprising result: \emph{the attention mechanisms in these models seem not only ineffective but also potentially detrimental}.

\begin{table}[t]
\setlength\tabcolsep{10pt}
\centering
\footnotesize
\caption{
The best Recall@10 achieved by each group of KGRSs at different sparsity levels on the Last.FM. Bold values highlight the top performance for each sparsity level. The "Improve" shows the relative improvement of LightKG over the best GNN-based KGRSs.}
\label{sparse_compare}%
\vspace{-0.15in}
{
\begin{tabular}{ccccc}
\toprule
Model      & 80\%              & 40\%            & 20\%              & 10\%              \\ \midrule
Path$_{max}$ & 0.2132           & 0.1243          & 0.0935           & 0.0825           \\
Embedding$_{max}$  & 0.2453           & 0.1797         & \textbf{0.1082} & \textbf{0.0983} \\
GNN$_{max}$        & \textbf{0.2727} & \textbf{0.1802} & 0.0876          & 0.0425          \\ \midrule
LightKG    & 0.2929          & 0.2120         & 0.1012          & 0.0861           \\
Improve    & 8.52\%           & 17.66\%         & 15.49\%          & 102.40\%         \\ 
\bottomrule
\end{tabular}
}
\vspace{-0.4cm}
\end{table}%

\subsection{Impact of Interaction Sparsity}
We evaluate 12 recent and representative KGRSs on Last.FM and ML-1M datasets (detailed statistics are provided in Tab.~\ref{tab:Statistics of the datasets}), using Recall@10 and MRR@10 as evaluation metrics.
To simulate varying degrees of sparse interactions, we build datasets with varying sparsity levels by adjusting the sampling ratio of the full datasets.
Sampling ratios are set according to the original density levels of each dataset, with 10\% and 5\% representing sparse scenarios, and 80\%, in comparison, serves as relatively dense scenario.
Hyperparameters are tuned for all models across varying sparsity levels, including the number of GNN layers, learning rates, negative sampling ratio, and other key parameters.

Fig.~\ref{fig:sparse} presents the experimental results, with the x-axis representing sampling ratio (e.g., 40\% means 40\% of the full dataset is sampled for training).
To systematically evaluate the performance of KGRSs in sparse scenarios, we categorized them into three groups based on their frameworks~\cite{jiang2024diffkg}: path-based (MCRec, RippleNet), embedding-based (CFKG, CKE), and GNN-based (KGCN, KGNNLS, KGAT, KGIN, MCCLK, KGRec, DiffKG, CL-SDKG). 
Tab.~\ref{sparse_compare} summarizes the highest Recall achieved by each group at different sparsity levels on the Last.FM dataset.
For example, "Path$_{max}$ in 40\%" represents the highest Recall achieved by path-based KGRSs (RippleNet, MCRec) at a 40\% sampling ratio.

The results reveal several key observations. 
\textbf{First}, as expected, the accuracy of all KGRSs declines with increasing data sparsity as shown in Fig.~\ref{fig:sparse} and Tab.~\ref{sparse_compare}. 
\textbf{Second}, while GNN-based KGRSs demonstrate superior performance in dense scenarios (sampling ratios of 80\% and 40\%), their performance degrades more significantly in sparse scenarios (sampling ratios of 20\% and 10\%) even with SSL, as shown in Tab.~\ref{sparse_compare}.
Notably, attempts to mitigate this degradation through expanded GNN receptive fields prove ineffective. In fact, introducing additional GNN layers often brings in noise, which outweighs the potential benefits of additional information~\cite{zhou2020graph}.
\textbf{Lastly}, Fig.~\ref{fig:sparse} reveals that simpler models like KGCN and KGNN-LS frequently outperform more complex counterparts such as KGAT and MCCLK in sparse scenarios. 
This indicates a notable correlation between model complexity and accuracy in sparse scenarios, which will be further explored in the following subsections.

\begin{table}[t]
\setlength{\abovecaptionskip}{3pt}
\centering
\footnotesize
  \caption{Complexity analysis and Recall@10 on Last.FM (sampling Ratio: 10\%) and ML-1M (sampling Ratio: 5\%). Correlation refers to the Pearson correlation coefficient between complexity ranking and Recall.}
  \label{complexivity}%
\vspace{-0.00in}
\resizebox{0.975\linewidth}{!}{
    \begin{tabular}{cccc}
    \toprule
    Model & Time Complexity (ranked from low to high) & Last.FM & ML-1M \\
    \midrule
    KGCN(2019)  & $\mathcal{O}(d(|\mathcal{G}_{KG}|+|\mathcal{R}|\times|\mathcal{U}|+|\mathcal{I}|))$ & 0.0425 & 0.0575 \\
    KGNNLS(2019) & $\mathcal{O}(d(|\mathcal{G}_{KG}|+|\mathcal{V}|+|\mathcal{U}|\times|\mathcal{R}|))$ & 0.0378 & 0.0542 \\
    KGIN(2021)  & $\mathcal{O}(d|\mathcal{G}_{KG}| + d^2|\mathcal{G}_{UI}|)$ & 0.0289 & 0.0557 \\
    KGRec(2023) & $\mathcal{O}(d^2|\mathcal{G}_{KG}| + d|\mathcal{G}_{UI}|)$ & 0.0291 & 0.0398 \\
    DiffKG(2024) & $\mathcal{O}(d^2|\mathcal{G}_{KG}| + d|\mathcal{G}_{UI}|)$ & 0.0177 & 0.0357 \\
    CL-SDKG(2024)  & $\mathcal{O}(d^2(|\mathcal{G}_{KG}|+|\mathcal{U}|) + d|\mathcal{G}_{UI}|)$ & 0.0204 & 0.0445 \\
    KGAT(2019)  & $\mathcal{O}(d^2(|\mathcal{G}_{KG}|+|\mathcal{G}_{UI}|+|\mathcal{U}|+|\mathcal{I}|+|\mathcal{V}|))$ & 0.0202 & 0.0311 \\
    MCCLK(2022) & $\mathcal{O}(d^3|\mathcal{G}_{KG}|+d|\mathcal{G}_{UI}|)$ & 0.0149 & 0.0475 \\
    \midrule
    Correlation &       & -0.9374 & -0.6682 \\
    \bottomrule
    \end{tabular}%
    }
\vspace{-0.4cm}
\end{table}%

\subsection{Impact of Model Complexity}
Due to the significant differences in frameworks among various types of KGRSs, comparing their complexities is challenging.
Therefore, we focus on GNN-based KGRSs, which is the most widely adopted framework~\cite{li2023survey}. 
We analyze the time complexity of a single GNN layer, as it serves as a fundamental building block for most GNN-based KGRSs.
Let $d$ denote the embedding size, while $|\mathcal{G}_{UI}|$ and $|\mathcal{G}_{KG}|$ represent the number of triplets in the user-item interaction graph and KG, respectively. 

Tab.~\ref{complexivity} ranks KGRSs by their time complexity and reports their Recall@10 on Last.FM (sampling ratio: 10\%) and ML-1M (sampling ratio: 5\%).
The correlation shows the Pearson correlation coefficient between complexity ranking and Recall@10, with negative values indicating an inverse relationship.
The results reveal the following observations. 
\textbf{(1) Impact of Complexity in Sparse Scenarios}: Models with higher complexity generally perform worse in sparse scenarios. 
This is supported by the strongly negative Pearson correlation between model complexity and Recall in sparse scenarios (e.g., -0.9374 on Last.FM and -0.6682 on ML-1M).
\textbf{(2) Effectiveness of SSL}: SSL helps mitigate the issue of insufficient training data in sparse scenarios by extracting more supervisory signals. This can be supported by the fact that although KGRec is more complex than KGIN, its SSL training mechanism leads to superior performance on Last.FM compared to KGIN.

\begin{table}[t]
\setlength\tabcolsep{12pt}
\centering
\footnotesize
\caption{The Recall@10 after removing the attention mechanism on Last.FM. "Average" and "Average$_{a-}$" represent the average Recall of the six SOTA KGRSs with and without attention mechanisms, respectively, while "Improve" indicates the latter's improvement over the former.
}
    \label{tab:attention_moved}
\vspace{-0.15in}
{
    \begin{tabular}{ccccc}
    \toprule
    Model & 80\%   & 40\%  & 20\%   & 10\% \\
    \midrule
    \multicolumn{1}{c}{KGAT} & 0.2583  & 0.1423  & \textbf{0.0677}  & 0.0202  \\
    \multicolumn{1}{c}{KGAT$_{a-}$} & \textbf{0.2690}  & \textbf{0.1466}  & 0.0667  & \textbf{0.0228}  \\
    \midrule
    \multicolumn{1}{c}{KGIN} & \textbf{0.2727}  & 0.1711  & 0.0770  & 0.0289  \\
    \multicolumn{1}{c}{KGIN$_{a-}$} & 0.2718  & \textbf{0.1718}  & \textbf{0.0796}  & \textbf{0.0307}  \\
    \midrule
    \multicolumn{1}{c}{KGRec} & 0.2560  & \textbf{0.1758}  & \textbf{0.0876}  & 0.0291  \\
    \multicolumn{1}{c}{KGRec$_{a-}$} & \textbf{0.2573}  & 0.1740  & 0.0872  & 0.0291  \\
    \midrule
    \multicolumn{1}{c}{MCCLK} & \textbf{0.2699}  & 0.1758  & 0.0555  & 0.0149  \\
    \multicolumn{1}{c}{MCCLK$_{a-}$} & 0.2698  & \textbf{0.1792}  & \textbf{0.0557}  & \textbf{0.0152}  \\
    \midrule
    \multicolumn{1}{c}{DiffKG} & 0.2520  & 0.1551  & 0.0479  & 0.0177  \\
    \multicolumn{1}{c}{DiffKG$_{a-}$} & \textbf{0.2564}  & \textbf{0.1599}  & \textbf{0.0488}  & \textbf{0.0214}  \\
    \midrule
    \multicolumn{1}{c}{CL-SDKG} & 0.2409  & \textbf{0.1591}  & 0.0431  & 0.0203  \\
    \multicolumn{1}{c}{CL-SDKG$_{a-}$} & \textbf{0.2424}  & 0.1571  & \textbf{0.0450}  & \textbf{0.0204}  \\
    \midrule
    Average & 0.2583  & 0.1632  & 0.0631  & 0.0219  \\
    Average$_{a-}$ & 0.2611  & 0.1648  & 0.0638  & 0.0233  \\
    \textbf{Improve} & \textbf{+1.09\% } & \textbf{+0.97\% } & \textbf{+1.10\% } & \textbf{+6.05\% } \\
    \bottomrule
    \end{tabular}%
}
\vspace{-0.4cm}
\end{table}%

\subsection{Impact of Removing Attention Mechanisms} 
Tab.~\ref{complexivity} shows that complex models may sometimes underperform simpler models in sparse scenarios. 
To further explore this issue, a natural approach is to examine whether and how making models simpler, e.g., removing their attention mechanisms, would influence recommendation accuracy.\footnote{A unified ablation study is challenging due to the structural and functional diversity of these models, and our focus is solely on verifying the influence of simpler models.}
Specifically, We implement it by fixing attention weights to 1 in six SOTA KGRSs and conduct the same experiment as in Section 3.1.
Notably, this removal is rather simple and can lead to substantial disruption of the model's structure. 
For example, the original structure of KGAT is as follows:
\begin{equation}
\footnotesize
a(h,r,t) = (\boldsymbol{W}_r\boldsymbol{e}_t)^\top\tanh\biggl((\boldsymbol{W}_r\boldsymbol{e}_h+\boldsymbol{e}_r)\biggr),
\end{equation}
\begin{equation}
\footnotesize
\pi(h,r,t) = \frac{\exp(a(h,r,t))}{\sum\nolimits_{(h,r',t')\in\mathcal{N}_h}\exp(a(h,r',t'))}, 
\end{equation}
\begin{equation}
\footnotesize
\boldsymbol{e}_{h}^{(l)} = \sum\nolimits_{(h,r,t)\in\mathcal{N}_h}\pi(h,r,t)\boldsymbol{e}_t^{(l-1)},
\end{equation}
where $\boldsymbol{W}_r \in \mathbb{R}^{d \times d}$ is the transformation matrix of relation $r$.
The removal of attention impairs KGAT's ability to account for the influence of different relations, simplifying its GNN structure as:
\begin{equation}
\footnotesize
\boldsymbol{e}_{h}^{(l)}=\sum\nolimits_{(h,r,t)\in\mathcal{N}_h}\boldsymbol{e}_t^{(l-1)}.
\end{equation}

The results are presented in Tab.~\ref{tab:attention_moved}.
The subscript $_{a-}$ (e.g., KGAT$_{a-}$) denotes the altered model (of KGAT) with the attention mechanism removed.
Due to space limitation, we only present the Recall@10 results for Last.FM, as similar conclusions are observed across other datasets in appendix.
Surprisingly, despite the simplicity of the removal method, we observed that: 
\textbf{(1)} \emph{in most cases, removing the attention mechanism slightly improves recommendation accuracy};
and \textbf{(2)} \emph{as the data sparsity level increases, the improvement becomes more pronounced}. 
These observations strongly support simpler models are more suited for sparse scenarios.

\textbf{Intuitively, complex models often require substantial training data to effectively learn their parameters and achieve optimal performance.
In sparse scenarios, they often underfit or generalize poorly.
Therefore, simpler models are better suited to address the challenges posed by sparse scenarios.}

\section{LightKG}
\begin{figure*}[htp]
    \centering  \includegraphics[width=0.85\textwidth]{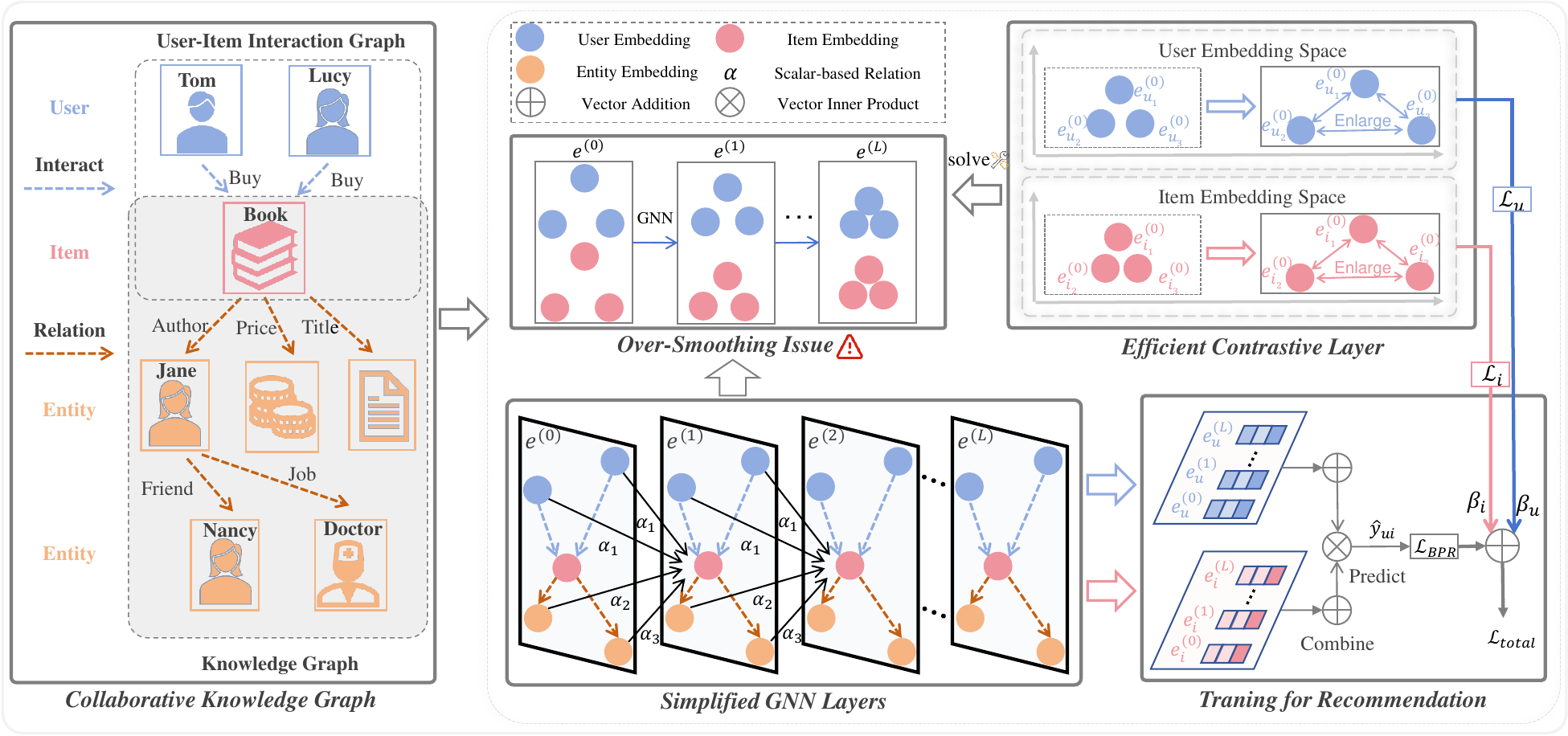}
    \vspace{-0.4cm}
    \caption{Illustration of our proposed LightKG.}
    \label{fig:framework}
    \vspace{-0.5cm}
\end{figure*}

Our previous experimental explorations inspire us to design a simpler yet effective KGRS, LightKG. 
Fig.~\ref{fig:framework} displays its workflow.
First, it employs a simplified GNN layer, with \emph{scalar-based relation encoding and linear aggregation framework}, for aggregating neighbor information. 
Second, to shorten the training time for SSL, it employs an efficient contrastive layer that directly minimizes the similarity between nodes of the same type, eliminating the need for generating and comparing subgraphs—a common yet computationally expensive approach adopted by previous SSL methods.
Finally, it derives matching scores for the recommendation task.

\subsection{Simplified GNN Layer}
We propose a simplified GNN layer, which serves to derive node representations in the CKGs.
The existing models such as KGRec, MCCLK, and KGAT encode relations as vectors or matrices, which are then incorporated into mechanisms such as attention and multi-graph learning.
To simplify the process of GNNs, we encode relations of the same type as scalar pairs, significantly reducing complexity while retaining essential information.
For example, consider the triplet (\textit{book1}, book-author, \textit{Jane}).
LightKG will encode the relation "book-author" as a pair of learned scalars to denote the importance of authors to books (''\textit{write}") and books to authors (''\textit{written by}"), respectively.
These scalars, similar to embeddings, are learned through backpropagation after initialization.
Although this may seem simplistic in capturing the semantic information in the CKGs, our subsequent experiments prove its effectiveness (refer to Tab.~\ref{fig:nokg}).

Inspired by the success of LightGCN~\cite{he2020lightgcn}, we propose a simple and linear GNN framework tailored for CKGs by leveraging scalar-based relations and excluding non-linear activation function. 
Specifically, for a node $k$, we use Equ.~\ref{lightkg} to update its embedding $\boldsymbol{e}_{k}^{(l)}$ in the $l$-th GNN layer,
\begin{equation}
\footnotesize
\label{lightkg}
\boldsymbol{e}_{k}^{(l)} = \sum\nolimits_{(k,r,t) \in \mathcal{N}_{k}} \frac{\alpha_{r_{tk}}}{\sqrt{|\mathcal{N}_{k}|} \sqrt{|\mathcal{N}_{t}}|} \boldsymbol{e}_{t}^{(l-1)},
\end{equation}
where $\alpha_{r_{tk}}$ denotes scalar-based relation from node $t$ to $k$. Note that $\alpha_{r_{tk}} \neq \alpha_{r_{kt}}$.
For each node of user ($u$) and item ($i$) in CKG, their final embeddings, which are used for model prediction, are generated by combining the outputs from each GNN layer:
\begin{equation}
\footnotesize
\boldsymbol{e}_{u}^{*} =\sum\nolimits_{l=0}^{L} \frac{1}{L+1} \boldsymbol{e}_{u}^{(l)},\boldsymbol{e}_{i}^{*} =\sum\nolimits_{l=0}^{L} \frac{1}{L+1} \boldsymbol{e}_{i}^{(l)}.
\end{equation}

By employing a simple and linear neighbor-information aggregation framework with scalar-based relation encoding, LightKG greatly simplifies the existing GNN-KGRSs, thereby reducing the learning difficulty especially in sparse scenarios.

\subsection{Efficient Contrastive Layer}
Contrastive learning, as a form of SSL, can alleviate the issue of insufficient supervised training data in sparse scenarios~\cite{yang2022knowledge}.
The recent advances in contrastive learning show that SSL is effective as it enhances the distinction between node embeddings~\cite{zou2022multi}.
Such distinction can help alleviate the over-smoothing issue~\cite{you2020graph}, a phenomenon where node embeddings in GNNs become indistinguishable due to excessive aggregation of neighborhood information.
However, existing contrastive learning methods, such as those used in MCCLK and KGRec, are time-consuming since they need to generate subgraphs and make comparisons between them.

To address it, we aim to simplify the contrastive learning with a specific focus on the over-smoothing issue.
Inspired by~\cite{wang2020understanding}, we propose to gain more distinctive node embeddings by directly minimizing similarities between embeddings of different nodes.
Empirically, we find this works best when applied separately to nodes of the same type, such as between users or between items, as follows,
\begin{equation}
\footnotesize
\label{eq:contrastive1}
\min \mathcal{L}_u = \sum\nolimits_{u_1, u_2 \in \mathcal{U}}{\exp(\boldsymbol{e}_{u_1}^{(0)\top} \boldsymbol{e}_{u_2}^{(0)})}, 
\end{equation}
\begin{equation}
\footnotesize
\label{eq:contrastive2}
\min \mathcal{L}_i = \sum\nolimits_{i_1, i_2 \in \mathcal{I}}{\exp(\boldsymbol{e}_{i_1}^{(0)\top} \boldsymbol{e}_{i_2}^{(0)})}, 
\end{equation}
where $\boldsymbol{e}_{u_1}^{(0)},\boldsymbol{e}_{u_2}^{(0)},\boldsymbol{e}_{i_1}^{(0)}$ and $ \boldsymbol{e}_{i_2}^{(0)}$ are normalized user and item embeddings at layer 0. We empirically find that adding the contrastive layer at layer 0 yields the best performance, as it enables the model to learn discriminative features from the outset, making deeper layer representations naturally distinct.

However, Equ.~\ref{eq:contrastive1} and Equ.~\ref{eq:contrastive2} ignore two key factors. 
First, they overlook the inherent similarity that exists between nodes. 
For two users who have highly overlapping interaction records, their embeddings should not be excessively differentiated. 
Second, over-smoothing impacts nodes unevenly, as those with more neighbors are more susceptible to losing their unique characteristics in GNNs.
Thus, it is crucial to consider the number of neighbors. 
By incorporating these factors, we refine Equ.~\ref{eq:contrastive1} and Equ.~\ref{eq:contrastive2} as follows:
\begin{equation}
\footnotesize
\label{eq:L_u}
\min \mathcal{L}_u = \sum\nolimits_{u_1, u_2 \in \mathcal{U}}{w_{u_1,u_2} \exp((1-s_{u_1,u_2})\boldsymbol{e}_{u_1}^{(0)\top} \boldsymbol{e}_{u_2}^{(0)})}, 
\end{equation}
\begin{equation}
\footnotesize
\label{eq:L_i}
\min \mathcal{L}_i = \sum\nolimits_{i_1, i_2 \in \mathcal{I}}{w_{i_1,i_2} \exp((1-s_{i_1,i_2})\boldsymbol{e}_{i_1}^{(0)\top} \boldsymbol{e}_{i_2}^{(0)})},
\end{equation}
where $s_{i,j} = \frac{\mathcal{N}_i \cap \mathcal{N}_j}{\sqrt{|\mathcal{N}_i| |\mathcal{N}_j|}}$ denotes the similarity, and $w_{i,j}=1-\frac{1}{\sqrt{|\mathcal{N}_i| |\mathcal{N}_j|}}$ reflects the impact of neighbor counts.
By optimizing Equ.~\ref{eq:L_u} and Equ.~\ref{eq:L_i}, we effectively and efficiently address the over-smoothing issue. Unlike prior works, we avoids the computationally expensive task of generating and comparing subgraphs. Consequently, our method not only enhances recommendation accuracy (see Fig.~\ref{fig:ablation}) but also greatly shortens the training time (see Tab.~\ref{Efficiency_Comparison}).

\subsection{Model Prediction and Loss Function}
The embeddings obtained from the GNN layer can be used to derive matching scores between users and items using $\hat{y}_{u i}=\boldsymbol{e}_{u}^{* \top} \boldsymbol{e}_{i}^{*}$.
For the recommendation task, we choose the BPR loss~\cite{rendle2012bpr}. 
It optimizes the personal ranking of items for each user by maximizing the probability that he/she prefers a positively interacted item $\hat{y}_{u i}$ over a randomly chosen non-interacted one $\hat{y}_{u j:}$
\begin{equation}
\footnotesize
\mathcal{L}_{\mathrm{BPR}}=\sum\nolimits_{i\in \mathcal{N}_u, \notin j\mathcal{N}_u}-\ln \sigma\left(\hat{y}_{u i}-\hat{y}_{u j}\right),
\end{equation}
where $\sigma$ is the sigmoid function. 
Together with our objectives for the contrastive layer, we obtain the following objective function:
\begin{equation}
\footnotesize
\label{eq:overall-loss}
\mathcal{L}_{total}=\mathcal{L}_{\mathrm{BPR}}+\beta_u \mathcal{L}_u+\beta_i \mathcal{L}_i +\lambda \|\Theta\|_2^2,
\end{equation}
where $\Theta$ represents the model parameter set; 
$\beta_u$ and $\beta_i$ are two hyperparameters that determine the respective strengths of $\mathcal{L}_u$ and $\mathcal{L}_i$; and 
$\lambda$ is the regularization coefficient. 

\subsection{Model Analysis}
In this section, we analyze the complexity of LightKG, showing that its time complexity is lower than that of all existing GNN-based KGRSs.
Therefore, LightKG is a simpler yet effective model, capable of achieving robust learning even in sparse scenarios.
We further demonstrate that LightKG can be seen as an extension of LightGCN, which means that LightKG inherits LightGCN’s efficient capability to extract information from interaction data. 
Finally, we discuss the encoding of relations as scalar pairs, which can be interpreted as node labels, enabling the model to distinguish different node types during representation learning.

\subsubsection{Complexity Analysis}
By encoding relations as scalar pairs and incorporating a linear aggregation framework, LightKG achieves a time complexity of $\mathcal{O}(d(|\mathcal{G}_{UI}|+|\mathcal{G}_{KG}|))$, which is significantly lower than that of existing GNN-based KGRSs. 
This reduced complexity not only improves computational efficiency but also effectively addresses learning challenges in sparse scenarios.
The simplicity of LightKG's architecture allows it to learn effectively from limited training data, making it particularly well suited for sparse scenarios, where traditional GNN-based KGRSs often struggle due to their complexity and high data demands.
Moreover, in dense scenarios, LightKG retains the strengths of GNN, leveraging rich interaction data to deliver stronger performance. 
Thus, LightKG is adaptable and effective for both sparse (see Tab.~\ref{sparse_compare}) and dense (see Tab.~\ref{tab:overall_performance}) recommendation scenarios.

\subsubsection{Relationship with LightGCN}
For a better comparison, here we consider only the interaction graphs, as LightGCN cannot be applied to KG {directly}.
We use $ \boldsymbol{A} \in \mathbb{R}^{\left|\mathcal{U}\right| \times \left|\mathcal{I}\right|}$ to denote the interaction matrix, where $\boldsymbol{A}_{(u,i)} = 1$ means an observed interaction between user $u$ and item $i$.
Based on this, the matrix form of GNN in LightGCN can be denoted as follows:
\begin{equation}
\footnotesize
\boldsymbol{E}^{\left(l\right)} = \boldsymbol{D}^{ - \frac{1}{2}} \begin{pmatrix} 0 & \boldsymbol{A} \\ \boldsymbol{A}^{T} & 0 \\  \end{pmatrix} \boldsymbol{D}^{ - \frac{1}{2}}\boldsymbol{E}^{\left(l-1\right)},
\end{equation}
where $\boldsymbol{D} \in \mathbb{R}^{(|\mathcal{U}|+|\mathcal{I}|) \times (|\mathcal{U}|+|\mathcal{I}|)}$ is a diagonal matrix.
For LightKG, we use a pair of scalars $\alpha_{u}$ and $\alpha_{i}$ to denote the relation encoding between users and items.
Then, the matrix form of GNN in LightKG can be denoted as follows:
\begin{equation}
\footnotesize
\boldsymbol{E}^{\left(l\right)} = \boldsymbol{D}^{ - \frac{1}{2}} \begin{pmatrix} 0 & \alpha_{iu}\boldsymbol{A} \\ \alpha_{ui}\boldsymbol{A}^{T} & 0 \\  \end{pmatrix} \boldsymbol{D}^{ - \frac{1}{2}}\boldsymbol{E}^{\left(l-1\right)}, \label{eq:1}
\end{equation}
\begin{equation}
\footnotesize
\begin{split}
\boldsymbol{E}^{\left(l\right)} &= \boldsymbol{D}^{ - \frac{1}{2}} \begin{pmatrix} 0 & \boldsymbol{A} \\ \boldsymbol{A}^T & 0 \\  \end{pmatrix} \boldsymbol{D}^{ - \frac{1}{2}}\boldsymbol{E}^{\left(l-1\right)} \\
&\quad + \boldsymbol{D}^{ - \frac{1}{2}} \begin{pmatrix} 0 & (\alpha_{iu}-1)\boldsymbol{A} \\ (\alpha_{ui}-1)\boldsymbol{A}^{T} & 0 \\  \end{pmatrix} \boldsymbol{D}^{ - \frac{1}{2}}\boldsymbol{E}^{\left(l-1\right)}. \label{eq:2}
\end{split}
\end{equation}
Since $\alpha_{iu}$ and $\alpha_{ui}$ are learnable, LightKG becomes equivalent to LightGCN when $\alpha_{iu} = \alpha_{ui} = 1$.
Therefore, LightKG can be viewed as an extension of LightGCN, inheriting its ability to efficiently extract information from interaction graph.
This is particularly important, as many items, especially new ones, may not link to KGs. 
Our subsequent experiments (Tab.~\ref{fig:nokg}) reveal that many KGRSs fail to outperform LightGCN when KGs are removed. In contrast, LightKG achieves superior accuracy both with and without KGs.

\subsubsection{Scalar-based Relations as Node Labels}
In LightKG, the relations are encoded as scalar pairs such as $\alpha_{iu}$ and $\alpha_{ui}$, which can be interpreted as implicit labels for different node types.
To clarify this concept clearly, we focus on interaction graphs for simplicity and without loss of generality. 
Considering an extreme scenario where all node embeddings ($\boldsymbol{e}$) are identical, and each node has the same number of neighbors ($|\mathcal{N}|$). 
In traditional models like LightGCN and KGAT, this uniformity leads to identical aggregation coefficients. 
As a consequence, all nodes converge to identical embeddings after GNN propagation, thereby losing the distinction between different node types such as users and items.
In contrast, LightKG leverages scalar weights to differentiate node types.
Taking one layer of the GNN as an example, the updated embeddings of user ($u$) and item ($i$) are computed as follows:
\begin{equation}
\footnotesize
\boldsymbol{e}_{u}^{(l)} = \sum_{i \in \mathcal{N}_u} \frac{\alpha_{iu}}{\sqrt{|\mathcal{N}_u|} \sqrt{|\mathcal{N}_i|}} \boldsymbol{e}_i^{(l-1)} = \frac{\alpha_{iu}}{|\mathcal{N}|} \boldsymbol{e},
\end{equation}
\begin{equation}
\footnotesize
\boldsymbol{e}_{i}^{(l)} = \sum_{u \in \mathcal{N}_{i}} \frac{\alpha_{ui}}{\sqrt{|\mathcal{N}_i|} \sqrt{|\mathcal{N}_u|}} \boldsymbol{e}_u^{(l-1)} = \frac{ \alpha_{ui}}{|\mathcal{N}|} \boldsymbol{e}.
\end{equation}
The distinction between $\alpha_{iu}$ and $\alpha_{ui}$ ensures that the user and item embeddings diverge after the propagation of GNN. 
These scalars act as implicit labels.
For example, $\alpha_{iu}$ is the label of user nodes, as it only appears in the expression of $\boldsymbol{e}_u$.

Our subsequent experiments (Fig.~\ref{fig:parameter}) reveal that the optimal values of $\beta_u$ and $\beta_i$ (refer to Equ.~\ref{eq:overall-loss}) may differ significantly, suggesting that the embeddings of users and items follow distinct distributions.
As proven in~\cite{bei2024correlation}, labeling different node types enable the model to effectively account for the unique characteristics of each node type, thereby improving its capacity to capture patterns specific to users and items during the representation learning process.

\begin{table}[t]
\setlength\tabcolsep{2pt}
\centering
\footnotesize
\caption{Statistics of the datasets.}
\label{tab:Statistics of the datasets}
\vspace{-0.15in}
{
\begin{tabular}{l|cccccc}
\toprule
\multirow{2}{*}{\textbf{Statistics}} & \multicolumn{3}{c}{\textbf{User-Item Graph}} & \multicolumn{3}{c}{\textbf{Knowledge Graph}} \\
 & users & items & interactions & entities & relations & triplets \\ \midrule
Amazon-book (AMZ-B) & 20,347 & 4,230 & 234,323 & 12,230 & 21 & 46,522 \\
MovieLens-1M (ML-1M) & 6,039 & 3,499 & 573,637 & 77,799 & 51 & 378,151 \\
Book-Crossing (BX) & 11,018 & 9,059 & 24,644 & 77,904 & 27 & 151,500 \\
Last.FM & 1,873 & 3,847 & 21,173 & 9,367 & 62 & 15,518 \\ 
\bottomrule
\end{tabular}
}
\vspace{-0.5cm}
\end{table}

\section{Experiments}
\label{sec:exp}
We conduct extensive experiments to demonstrate the effectiveness and efficency of our proposed LightKG. We aim to answer the following four research questions:
\begin{itemize}[leftmargin=*]
\item RQ1: How does LightKG perform compared to the SOTA KGRSs, in terms of both the recommendation accuracy and the training efficiency in dense scenarios?
\item RQ2: How does LightKG perform in sparse scenarios compared to the SOTA KGRSs?
\item RQ3: With its simplified relation encodings and aggregation framework, does LightKG effectively utilize KGs in recommendation compared to the SOTA KGRSs?
\item RQ4: How does the contrastive layer contribute to the overall performance {of our proposed LightKG}? 
\end{itemize}

\subsection{Experimental Settings}
\subsubsection{Datasets and Evaluation Protocol}
 We select four benchmark datasets: Last.FM\footnote{https://grouplens.org/datasets/hetrec-2011/}, Amazon-Book\footnote{https://jmcauley.ucsd.edu/data/amazon/} {(AMZ-B)}, Book-Crossing\footnote{http://www2.informatik.uni-freiburg.de/~cziegler/BX/} (BX), and MovieLens-1M\footnote{https://grouplens.org/datasets/movielens/1m/} (ML-1M).
Following SOTA methods~\cite{li2022gromov}, we preprocess ML-1M and {AMZ-B} by retaining interactions with ratings of at least 4. For {AMZ-B}, we further apply a 20-core setting, ensuring that both users and items have a minimum of 20 interactions. For Last.FM and BX, we use the versions released in~\cite{wang2018ripplenet,wang2019KGCN} without additional modifications.
Tab.~\ref{tab:Statistics of the datasets} presents the statistical details of the datasets.
For evaluation, we use the full-rank approach to generate top-10 recommendations. 
To evaluate the recommendation accuracy, we adopt widely used metrics, Recall@10 and MRR@10, following~\cite{zhang2024does}.

\subsubsection{Baselines}
We compare our proposed LightKG with 13 SOTA baselines, including KG-free RS (LightGCN~\cite{he2020lightgcn}), embedding-based KGRSs (CFKG~\cite{zhang1803learning}, CKE~\cite{zhang2016collaborative}), path-based KGRSs (RippleNet~\cite{wang2018ripplenet}, MCRec~\cite{hu2018leveraging}), supervised GNN-based KGRSs (KGCN~\cite{wang2019KGCN}, KGNN-LS~\cite{wang2019knowledge}, KGAT~\cite{wang2019kgat}, KGIN~\cite{wang2021learning}) and self-supervised GNN-based KGRSs (MCCLK~\cite{zou2022multi}, KGRec~\cite{yang2023knowledge}, DiffKG~\cite{jiang2024diffkg}, CL-SDKG~\cite{shi2024self}).

\begin{table}[t]
\setlength{\abovecaptionskip}{6pt}
\setlength\tabcolsep{4pt}
\centering
\caption{The results of Recall@10 and MRR@10 of all methods {on the four benchmark datasets}. * denotes statistically significant
different by the paired t-test with $p-\text{value}< 0.01$.}
\label{tab:overall_performance}
\vspace{-0.05in}
\resizebox{0.49\textwidth}{!}{
\begin{tabular}{cllllllll}
    \toprule
    \multicolumn{1}{c}{\multirow{2}[2]{*}{Model}} & \multicolumn{2}{c}{\textbf{AMZ-B}} & \multicolumn{2}{c}{\textbf{ML-1M}} & \multicolumn{2}{c}{\textbf{BX}} & \multicolumn{2}{c}{\textbf{Last.FM}} \\
    \cmidrule(lr){2-3} \cmidrule(lr){4-5} \cmidrule(lr){6-7} \cmidrule(lr){8-9}
    \multicolumn{1}{c}{} & \multicolumn{1}{c}{\textbf{Recall}} & \multicolumn{1}{c}{\textbf{MRR}} & \multicolumn{1}{c}{\textbf{Recall}} & \multicolumn{1}{c}{\textbf{MRR}} & \multicolumn{1}{c}{\textbf{Recall}} & \multicolumn{1}{c}{\textbf{MRR}} & \multicolumn{1}{c}{\textbf{Recall}} & \multicolumn{1}{c}{\textbf{MRR}} \\
    \midrule
    LightGCN & 0.1980 & 0.1052 & 0.1855 & 0.3453 & 0.0468 & 0.0198 & 0.2695 & 0.1204 \\
    CFKG  & 0.1968 & 0.0987 & 0.1862 & 0.3405 & 0.0802 & 0.0384 & 0.2444 & 0.1100 \\
    CKE   & 0.1979 & 0.1037 & 0.1848 & 0.3457 & 0.0313 & 0.0152 & 0.2453 & 0.1069 \\
    RippleNet & 0.1561 & 0.0838 & 0.1590 & 0.3062 & 0.0445 & 0.0194 & 0.1633 & 0.0656 \\
    MCRec & 0.1524 & 0.0791 & 0.1610 & 0.3233 & 0.0512 & 0.0241 & 0.2132 & 0.0941 \\
    KGCN  & 0.1550 & 0.0738 & 0.1594 & 0.3456 & 0.0867 & 0.0433 & 0.2149 & 0.0924 \\
    KGNN-LS & 0.1508 & 0.0750 & 0.1592 & 0.3051 & 0.0731 & 0.0371 & 0.2117 & 0.0891 \\
    KGAT  & 0.1925 & 0.0997 & 0.1830 & 0.3412 & 0.0499 & 0.0233 & 0.2583 & 0.1152 \\
    KGIN  & \underline{0.2090} & 0.1099 & \underline{0.1969} & 0.3551 & 0.0801 & 0.0399 & \underline{0.2727} & \underline{0.1242} \\
    MCCLK & 0.2025 & 0.1065 & 0.1853 & 0.3474 & 0.0607 & 0.0359 & 0.2699 & 0.1228 \\
    KGRec & 0.2035 & 0.1094 & 0.1960 & \underline{0.3570} & \underline{0.1033} & \textbf{0.0540} & 0.2560 & 0.1118 \\
    DiffKG & 0.2039 & 0.1116 & 0.1846 & 0.3428 & 0.0581 & 0.0318 & 0.2520 & 0.1192 \\
    CL-SDKG & 0.2036 & \underline{0.1134} & 0.1861 & 0.3428 & 0.0924 & \underline{0.0532} & 0.2409 & 0.1054 \\
    \midrule
    LightKG & \textbf{0.2120*} & \textbf{0.1173*} & \textbf{0.2029*} & \textbf{0.3785*} & \textbf{0.1154*} & 0.0515* & \textbf{0.2929*} & \textbf{0.1350*} \\
    \bottomrule
    \end{tabular}%
}
\vspace{-0.4cm}
\end{table}
\subsubsection{Implementation Details}
All experiments are conducted on Ubuntu 18.04 with an Intel(R) Xeon(R) Gold 6226R CPU running at 2.90GHz, 64GB of memory, and 8 NVIDIA GeForce GTX 3090 GPU. 
To reduce randomness, all experiments are repeated five times.
To ensure the rigor of the experiments, we implement LightKG and all the baselines in RecBole~\cite{zhao2022recbole}, which is a unified, comprehensive and efficient recommendation library. 
We set the "stopping-step" parameter in Recbole to 20, which means that the model will stop training if no improvement is observed on the validation set for 20 consecutive epochs. 
The datasets are split into training, validation, and testing sets using a ratio of 8:1:1.
For a fair comparison, the embedding size of all the models is fixed to 64 and the Adam optimizer is used for optimization with a fixed batch size of 2048. 
We use Bayesian Optimization~\cite{sun2022daisyrec} for hyperparameter tuning and each model is optimized 30 trails per dataset.
All parameters including scalar-based relations are initialized by Xavier uniform.
Specifically, we search learning rate in $\{ 0.01, 0.005, 0.001, 0.0005, 0.0001\}$, the GNN layers in $\{1,2,3,4\}$ for all GNN-based KGRSs and the number of negative samples for training in $\{1,2,5,10\}$. 
For $\beta_u$ and $\beta_i$ (refer to Equ.~\ref{eq:overall-loss}), we tune them in $\{10^{-4},10^{-5},10^{-6},10^{-7},10^{-8},10^{-9}\}$.

\subsection{Experimental Results and Analysis}
\subsubsection{Result of RQ1.}
We evaluate both the recommendation accuracy and efficiency of LightKG in dense scenario (sampling ratio: 80\%).
The results are presented in Tab.~\ref{tab:overall_performance}, where the bolded numbers represent the best results and the underlined numbers indicate the second-best. 
We can draw the following conclusions.

\emph{First, while LightKG is specifically designed for sparse scenarios, it also demonstrates outstanding performance in dense scenarios}.
LightKG performs exceptionally well across all benchmark datasets, achieving the highest \textit{Recall/MRR} on Last.FM, AMZ-B and ML-1M, with improvements ranging from 1.42\% to 10.5\%.
For BX, it obtains the highest \textit{Recall} while ranking third in \textit{MRR}, closely following KGRec and CL-SDKG by a narrow margin.
These results highlight the strong generalizability of our simplified GNN framework.

\emph{Second, some KGRSs, like KGAT, perform worse than LightGCN which does not utilize KGs, highlighting their limitations in effectively using KGs or uncovering valuable insights from interaction graphs.}
Similar phenomenons are observed in \cite{yang2022knowledge}.
Note that LightGCN outperforms KGAT on AMZ-B, Last.FM and Ml-1M. 
It aligns with the slight improvement we observed after removing the attention mechanism from KGAT (as shown in Tab.~\ref{tab:attention_moved}), as KGAT becomes more similar to LightGCN once its attention mechanism is removed. 

\emph{Third, no single baseline model consistently outperforms the others across the four datasets.} 
Self-supervised methods (e.g., MCCLK and KGRec) do not always surpass supervised methods (e.g., KGIN). 
This may stem from the limitations of random graph augmentation or overly simplistic, handcrafted cross-view pairing, which may fail to capture meaningful KG information.

Lastly, we compare the training efficiency, i.e., time consumed per training epoch, between LightKG and other GNN-based KGRSs, as shown in Tab.~\ref{Efficiency_Comparison}. 
The results indicate that \emph{LightKG achieves a substantially shorter training time than other self-
supervised models, highlighting the efficiency of our designed contrastive layer.}
Although models like KGCN, KGNN-LS and KGAT achieve better training efficiency than LightKG due to the absence of SSL methods, they fail to obtain accurate recommendation.

\begin{table}[t]
\setlength{\abovecaptionskip}{3pt}
\centering
\footnotesize
\caption{Recall@10 on KG exploitation experiment. Bold values indicate the best Recall and improvement among all models and the underlined values indicate the second-best.
}
\label{fig:nokg}
\vspace{-1pt}
\resizebox{0.975\linewidth}{!}{
\begin{tabular}{ccccccc}
\toprule
     & \multicolumn{3}{c}{Last.FM} & \multicolumn{3}{c}{BX} \\
         & KG       & \text{w/o KG}  & Improve     & KG         & \text{w/o KG}     & Improve      \\ \midrule
LightGCN & --   & \underline{0.2695} & --  & --    & 0.0468  & --   \\
CFKG     & 0.2444  & 0.2389 & 2.29\%  & 0.0802     & \underline{0.0579}    & 38.51\%   \\
KGAT     & 0.2583  & 0.2640  & -2.17\% & 0.0499     & 0.0267    & 86.89\%   \\
KGIN     & \underline{0.2727}   & 0.2588 & \underline{5.37\%}  & 0.0801    & 0.0345    & \underline{132.12\%}   \\
MCCLK    & 0.2699   & 0.2608 & 3.49\%  & 0.0607    & 0.0306    & 98.30\%   \\
KGRec    & 0.2560    & 0.2571 & -0.43\% & \underline{0.1033}    & 0.0487    & 112.07\%   \\
DiffKG   & 0.2520    & 0.2517 & 0.12\%  & 0.0581     & 0.0210     & \textbf{176.67\%}   \\
SDKG     & 0.2327   & 0.2323 & 0.17\%  & 0.0924     & 0.0462    & 99.87\%  \\
\midrule
LightKG  & \textbf{0.2929}   & \textbf{0.2725} & \textbf{7.49\%}  & \textbf{0.1154}    & \textbf{0.0654}    & 76.48\% \\ \bottomrule
\end{tabular} }
\vspace{-0.4cm}
\end{table}

\subsubsection{Results of RQ2}
We now investigate the performance of LightKG across different sparsity levels.
Following the experimental setup in Section 4.1, we employ random sampling to generate datasets of varying sizes, representing scenarios with different sparsity levels. 
The specific sampling ratios used are $\{80\%, 40\%, 20\%, 10\%\}$ on Last.FM and $\{80\%, 40\%, 20\%, 10\%, 5\%\}$ on ML-1M, with each value denoting the proportion of interaction records allocated to the training set.
The results are presented in Tab.~\ref{sparse_compare} and Fig.~\ref{fig:sparse}.

\emph{\textbf{First}, when interaction records are sparse, LightKG outperforms all other GNN-based KGRSs. }
As sparsity level increases, LightKG's advantage over other GNN-based KGRSs becomes more evident, ranging from 8.52\% to 102.4\% on Last.FM. 
This supports our hypothesis that overly complex structures for capturing full KG semantics are not always beneficial and can hinder learning in sparse scenarios.
\emph{\textbf{Second}, despite outperforming other GNN-based KGRSs, LightKG falls behind CFKG, which is an embedding-based KGRS.}
We attribute this to the lower complexity of CFKG, which is only $\mathcal{O}(d \times batch\_size)$, making it even simpler than LightKG. 
Compared to CFKG, LightKG exhibits a clear advantage (with an improvement of 20.1\% on average) in dense scenarios and falls slightly behind (with a drop of 1.39\% on average) in sparse scenarios. 
Therefore, considering performance across both sparse and dense scenarios, LightKG shows optimal overall effectiveness.

\subsubsection{Results of RQ3.}
To check whether LightKG can capture the semantic information in KG with scalar-based relations, we remove the KG (denoted as \text{w/o KG}) and compare the performance before and after the removal, following the setting of~\cite{zhang2024does}.
We select several recently released or well-performing SOTA models for comparison.

\definecolor{airforceblue}{rgb}{0.36, 0.54, 0.66}
\definecolor{aliceblue}{rgb}{0.94, 0.97, 1.0}
\definecolor{alizarin}{rgb}{0.82, 0.1, 0.26}
\definecolor{almond}{rgb}{0.94, 0.87, 0.8}
\definecolor{amber}{rgb}{1.0, 0.75, 0.0}
\definecolor{amber(sae/ece)}{rgb}{1.0, 0.49, 0.0}
\definecolor{amethyst}{rgb}{0.6, 0.4, 0.8}
\definecolor{antiquebrass}{rgb}{0.8, 0.58, 0.46}
\definecolor{antiquefuchsia}{rgb}{0.57, 0.36, 0.51}
\definecolor{applegreen}{rgb}{0.55, 0.71, 0.0}
\definecolor{apricot}{rgb}{0.98, 0.81, 0.69}
\definecolor{arylideyellow}{rgb}{0.91, 0.84, 0.42}
\definecolor{ashgrey}{rgb}{0.7, 0.75, 0.71}
\definecolor{atomictangerine}{rgb}{1.0, 0.6, 0.4}
\definecolor{aureolin}{rgb}{0.99, 0.93, 0.0}
\definecolor{azure(colorwheel)}{rgb}{0.0, 0.5, 1.0}
\definecolor{babypink}{rgb}{0.96, 0.76, 0.76}
\definecolor{bluebell}{rgb}{0.64, 0.64, 0.82}
\definecolor{brightlavender}{rgb}{0.75, 0.58, 0.89}
\begin{figure}[t]
\centering
\hspace{-0.05in}
\subfigure{
\begin{tikzpicture}[scale=0.4]
\pgfplotsset{%
    width=0.4\textwidth,
    height=0.35\textwidth
}
\begin{axis}[
    ybar,
    bar width=6pt,
    ylabel={Recall@10},
    ylabel style ={font = \huge},
    xlabel style ={font = \normalsize},
    enlarge x limits={abs=0.7cm},
    scaled ticks=false,
    tick label style={/pgf/number format/fixed, font=\large},
    ymin=0.05, ymax=0.3,
    symbolic x coords={Last.FM, BX, ML-1M, AMZ-B},
    xtick=data,
    ytick={0.1, 0.2,0.3},
    legend style={at={(0.5,0.98)}, anchor=north,legend columns=4, column sep=0.15cm, draw=none, font=\huge},
]
\addplot [fill=almond] coordinates {
(Last.FM, 0.2981) (BX, 0.11482) (ML-1M, 0.2060) (AMZ-B, 0.2133)
};
\addplot [fill=amethyst] coordinates {
(Last.FM, 0.2932) (BX, 0.109) (ML-1M, 0.191) (AMZ-B, 0.2137)
};
\addplot [fill=applegreen] coordinates {
(Last.FM, 0.2955) (BX, 0.1116) (ML-1M, 0.201) (AMZ-B, 0.211)
};
\addplot [fill=babypink] coordinates {
(Last.FM, 0.1795) (BX, 0.0747) (ML-1M, 0.1821) (AMZ-B, 0.2034)
};
\end{axis}
\end{tikzpicture}
}
\hspace{0.2cm}
\subfigure{
\begin{tikzpicture}[scale=0.4]
\pgfplotsset{%
    width=0.4\textwidth,
    height=0.35\textwidth
}
\begin{axis}[
    ybar,
    bar width=6pt,
    ylabel={MRR@10},
    ylabel style ={font = \huge},
    xlabel style ={font = \huge},
    enlarge x limits={abs=0.7cm},
    scaled ticks=false,
    tick label style={/pgf/number format/fixed, font=\large},
    ymin=0.02, ymax=0.4,
    symbolic x coords={Last.FM, BX, ML-1M, AMZ-B},
    xtick=data,
    ytick={0.1, 0.2, 0.3,0.4},
    legend image post style={scale=2},
    legend style={at={(1.9,0.5)}, anchor=east,legend columns=1, row sep=0.5cm, draw=none, font=\huge},
]
\addplot [fill=almond] coordinates {
(Last.FM, 0.1369) (BX, 0.0513) (ML-1M, 0.3785) (AMZ-B, 0.1178)
};
\addplot [fill=amethyst] coordinates {
(Last.FM, 0.138) (BX, 0.0536) (ML-1M, 0.3446) (AMZ-B, 0.1183)
};
\addplot [fill=applegreen] coordinates {
(Last.FM, 0.1372) (BX, 0.0519) (ML-1M, 0.3602) (AMZ-B, 0.117)
};
\addplot [fill=babypink] coordinates {
(Last.FM, 0.0739) (BX, 0.0365) (ML-1M, 0.3353) (AMZ-B, 0.1124)
};
\legend{LightKG, $\text{LightKG}_{\text{w/o u}}$, $\text{LightKG}_{\text{w/o i}}$,$\text{LightKG}_{\text{w/o ui}}$}
\end{axis}
\end{tikzpicture}
}
\vspace{-0.15in}
\caption{The impacts of contrastive layer.}
      \label{fig:ablation}
\vspace{-0.5cm}
\end{figure}
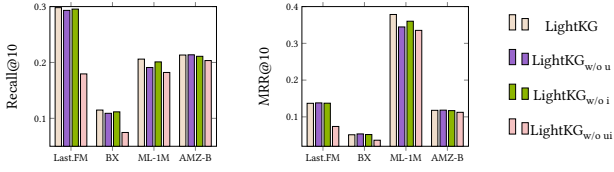

The results are illustrated in Tab.~\ref{fig:nokg}.
Notably, the "Improve" here ignores marginal effects of performance improvements.
For example, a 50\% increase from 60 to 90 is much harder than from 10 to 15.
\emph{\textbf{(1)} With the integration of the KG, LightKG shows a significant improvement in Recall, indicating its effective utilization of the KG to improve recommendation accuracy. }
This demonstrates that, despite the seemingly simplistic approach of encoding relations as scalar pairs, it effectively captures the semantic information of the KG while reducing model complexity.
\emph{\textbf{(2)} Many models perform worse than LightGCN when completely deprived of KG, indicating their limited ability to extract meaningful information solely from interaction graphs.} 
As demonstrated previously, LightKG, as an extension of LightGCN, can effectively extract meaningful information from interaction graphs.
Therefore, LightKG maintains high accuracy even with a reduced or absent KG, which demonstrates its robustness and adaptability in various scenarios.
\emph{\textbf{(3)} In certain scenarios, the removal of KGs leads to improved accuracy for some models.} 
For example, removing KG from KGAT and KGRec improves accuracy on Last.FM. 
This suggests that some overly complex framework may backfire, failing to effectively leverage KGs. 
The same phenomenons have also been observed in~\cite{zhang2024does}.

\subsubsection{Results of RQ4.}
We aim to evaluate the effectiveness of contrastive layers by conducting ablation experiments with three variants of LightKG. 
    \begin{itemize}[leftmargin=0.3cm]
\item $\textbf{LightKG}_{\text{w/o u}}$: This variant eliminates the contrastive layer for users by setting $\beta_u=0$.
\item  $\textbf{LightKG}_{\text{w/o i}}$: This variant eliminates the contrastive layer for items by setting $\beta_i=0$.
\item  $\textbf{LightKG}_{\text{w/o ui}}$: This variant eliminates contrastive layers for both users and items by setting $\beta_u = \beta_i=0$.
    \end{itemize}

The results in Fig.~\ref{fig:ablation} lead to the following observations. 
\textbf{(1)} \emph{Comparing LightKG and LightKG$_{\text{w/o ui}}$, we observe a notable drop in performance for LightKG$_{\text{w/o ui}}$ after the removal of the contrastive layers.} 
This highlights the critical role and effectiveness of contrastive layers, emphasizing their importance in enhancing the overall performance of LightKG. 
\textbf{(2)} \emph{A single contrastive layer (either user-side or item-side) can achieve performance comparable to the complete framework.}
For example, LightKG$_{\text{w/o u}}$, LightKG$_{\text{w/o i}}$ and LightKG achieve similar performance on Last.FM and BX.
This phenomenon occurs because the CKG connects users and items through a unified graph structure.
When the contrastive layer is applied to one side (e.g., users or items), the impact naturally spreads to the other side.

\definecolor{1}{RGB}{192, 0, 0}
\definecolor{2}{RGB}{0, 112, 192}
\definecolor{3}{RGB}{255,247,102}
\definecolor{4}{RGB}{255,178,102}
\definecolor{11}{RGB}{146,209,79}
\definecolor{5}{RGB}{194,209,163}
\definecolor{6}{RGB}{132,153,255}
\definecolor{7}{RGB}{163,223,196}
\definecolor{8}{RGB}{194,163,163}
\definecolor{9}{RGB}{194,102,163}
\definecolor{10}{RGB}{255,102,102}
\definecolor{red1}{RGB}{199,21,133}
\definecolor{red2}{RGB}{255,102,102}
\definecolor{red3}{RGB}{255,182,193}
\definecolor{blue1}{RGB}{25,25,112}
\definecolor{blue2}{RGB}{30,144,255}
\definecolor{blue3}{RGB}{135,206,250}
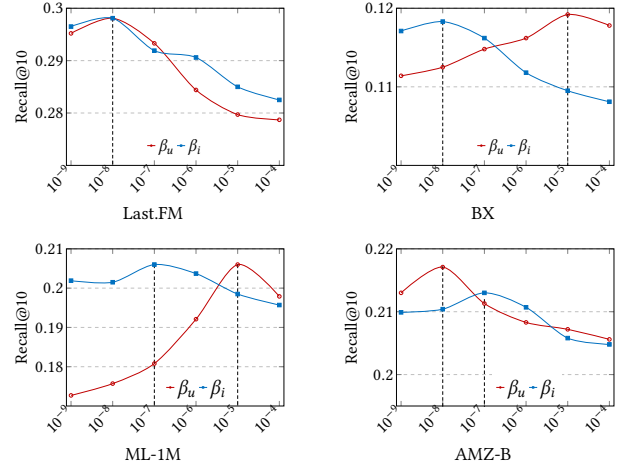
\begin{figure}[t]
	\centering
	\footnotesize
\subfigure[Last.FM]{
		\begin{tikzpicture}[scale=0.45]
		\begin{axis}[
            width=0.45\textwidth, height=0.35\textwidth,
	    ylabel={\huge Recall@10},
		xmin=0.99, xmax=6.01,
		ymin=0.27, ymax=0.3,
		xtick={1,2,3,4,5,6},
            xticklabels={$10^{-9}$,$10^{-8}$,$10^{-7}$,$10^{-6}$,$10^{-5}$,$10^{-4}$},
		yticklabel style={/pgf/number format/.cd,fixed,precision=3},
		ytick={0.28,0.29,0.3},
		ylabel style ={font = \huge},
        x tick label style={rotate=45, anchor=east}, 
		xlabel style ={font = \huge}, enlarge x limits= 0.02,
		scaled ticks=false,
        ylabel near ticks, 
		tick label style={font = \huge},             legend image post style={scale=0.4},
		legend style={at={(0.5,0.2)},font = \huge, anchor=north,legend columns=2,draw=none},
		ymajorgrids=true,
		grid style=dashed,
            mark size = 1.5pt,
		]
		\addplot[color=1,
		mark=o,
		line width=0.8pt,
		smooth]coordinates {
			( 1 , 0.2952 )
            ( 2, 0.2981 )
            ( 3, 0.2933 )
            ( 4 , 0.2844 )
            (5, 0.2797)
            (6, 0.2787)
		};
		\addplot[color=2,
		mark=square*,
		line width=0.8pt,
		smooth] coordinates {
        ( 1 , 0.2965 )
            ( 2, 0.2981 )
            ( 3, 0.2919 )
            ( 4 , 0.2906 )
            (5, 0.285)
            (6, 0.2825)
			};	
    \node[above] at (axis cs:2, 0.2981){}; 
    \draw[densely dashed] (axis cs:2, 0.2981) -- (axis cs:2,0); 
		\legend{$\beta_u$,$\beta_i$}
		\end{axis}
		\end{tikzpicture}
  }
\hspace{0.5cm}
\subfigure[BX]{
        \begin{tikzpicture}[scale=0.45]
		\begin{axis}[
            width=0.45\textwidth, height=0.35\textwidth,
	    ylabel={\huge Recall@10},
		xmin=0.99, xmax=6.01,
		ymin=0.1, ymax=0.12,
		xtick={1,2,3,4,5,6},
            xticklabels={$10^{-9}$,$10^{-8}$,$10^{-7}$,$10^{-6}$,$10^{-5}$,$10^{-4}$},
		yticklabel style={/pgf/number format/.cd,fixed,precision=3},
		ytick={0.11,0.12},
		ylabel style ={font = \huge},
        x tick label style={rotate=45, anchor=east}, 
		xlabel style ={font = \huge}, enlarge x limits= 0.02,
		scaled ticks=false,
        ylabel near ticks, 
		tick label style={font = \huge},             legend image post style={scale=0.4},
		legend style={at={(0.5,0.2)},font = \huge, anchor=north,legend columns=2,draw=none},
		ymajorgrids=true,
		grid style=dashed,
            mark size = 1.5pt,
		]
            
		\addplot[color=1,
		mark=o,
		line width=0.8pt,
		smooth]coordinates {
		( 1 , 0.1114 )
            ( 2 , 0.1125 )
            ( 3 , 0.1148 )
            ( 4 , 0.1162 )
            (5, 0.1192)
            (6, 0.1178)
		};
            
		\addplot[color=2,
		mark=square*,
		line width=0.8pt,
		smooth] coordinates {
            ( 1 , 0.1171 )
            ( 2 , 0.1183 )
            ( 3 , 0.1162 )
            ( 4 , 0.1118 )
            (5, 0.1095)
            (6, 0.1081)
			};	
            \draw[densely dashed] (axis cs:5, 0.1192) -- (axis cs:5,0); 
            \draw[densely dashed] (axis cs:2 , 0.1183) -- (axis cs:2,0); 
            
		\legend{$\beta_u$,$\beta_i$}
		\end{axis}
		\end{tikzpicture}
}
\vspace{-0.1cm}

\subfigure[ML-1M]{
        \begin{tikzpicture}[scale=0.45]
		\begin{axis}[
            width=0.45\textwidth, height=0.35\textwidth,
	    ylabel={\huge Recall@10},
		xmin=0.99, xmax=6.01,
		ymin=0.17, ymax=0.21,
		xtick={1,2,3,4,5,6},
            xticklabels={$10^{-9}$,$10^{-8}$,$10^{-7}$,$10^{-6}$,$10^{-5}$,$10^{-4}$},
		yticklabel style={/pgf/number format/.cd,fixed,precision=3},
		ytick={0.18,0.19,0.20,0.21},
		ylabel style ={font = \huge},
        x tick label style={rotate=45, anchor=east}, 
		xlabel style ={font = \huge}, enlarge x limits= 0.02,
		scaled ticks=false,
        ylabel near ticks, 
		tick label style={font = \huge},             legend image post style={scale=0.4},
		legend style={at={(0.6,0.2)},font = \Huge, anchor=north,legend columns=2,draw=none},
		ymajorgrids=true,
		grid style=dashed,
            mark size = 1.5pt,
		]
		\addplot[color=1,
		mark=o,
		line width=0.8pt,
		smooth]coordinates {
			( 1 , 0.1727 )
            ( 2 , 0.1757 )
            ( 3 , 0.1808 )
            ( 4 , 0.1921 )
            (5, 0.2060)
            (6, 0.1979)
		};
            
		\addplot[color=2,
		mark=square*,
		line width=0.8pt,
		smooth] coordinates {
			( 1 , 0.2019 )
            ( 2 , 0.2015 )
            ( 3 , 0.206 )
            ( 4 , 0.2037 )
            (5, 0.1985)
            (6, 0.1957)
			};	
            \draw[densely dashed] (axis cs:5, 0.2060) -- (axis cs:5,0); 
            \draw[densely dashed] (axis cs:3 , 0.206) -- (axis cs:3,0); 
            
		\legend{$\beta_u$,$\beta_i$}
		\end{axis}
		\end{tikzpicture}
}
\hspace{0.5cm}
\subfigure[AMZ-B]{
        \begin{tikzpicture}[scale=0.45]
		\begin{axis}[
            width=0.45\textwidth, height=0.35\textwidth,
	    ylabel={\huge Recall@10},
		xmin=0.99, xmax=6.01,
		ymin=0.195, ymax=0.22,
		xtick={1,2,3,4,5,6},
            xticklabels={$10^{-9}$,$10^{-8}$,$10^{-7}$,$10^{-6}$,$10^{-5}$,$10^{-4}$},
		yticklabel style={/pgf/number format/.cd,fixed,precision=3},
		ytick={0.20,0.21,0.22},
		ylabel style ={font = \huge},
        x tick label style={rotate=45, anchor=east}, 
		xlabel style ={font = \huge}, enlarge x limits= 0.02,
		scaled ticks=false,
        ylabel near ticks, 
		tick label style={font = \huge},             legend image post style={scale=0.4},
		legend style={at={(0.6,0.2)},font = \Huge, anchor=north,legend columns=2,draw=none},
		ymajorgrids=true,
		grid style=dashed,
            mark size = 1.5pt,
		]
		\addplot[color=1,
		mark=o,
		line width=0.8pt,
		smooth]coordinates {
			( 1 , 0.213 )
            ( 2 , 0.2171 )
            ( 3 , 0.2113 )
            ( 4 , 0.2083 )
            (5, 0.2072)
            (6, 0.2056)
		};
            
		\addplot[color=2,
		mark=square*,
		line width=0.8pt,
		smooth] coordinates {
			( 1 , 0.2099 )
            ( 2 , 0.2104 )
            ( 3 , 0.213 )
            ( 4 , 0.2107 )
            (5, 0.2058)
            (6, 0.2048)
			};	
            \draw[densely dashed] (axis cs:2, 0.2172) -- (axis cs:2,0); 
            \draw[densely dashed] (axis cs:3 , 0.212) -- (axis cs:3,0); 
            
		\legend{$\beta_u$,$\beta_i$}
		\end{axis}
		\end{tikzpicture}
}
 	\vspace{-0.15in}
	\caption{The impacts of $\beta_u$ and $\beta_i$.}
	\label{fig:parameter}
 	\vspace{-0.5cm}
\end{figure}

To further explore the impact of contrastive layer parameters, we fixed the parameter $\beta_u$ and $\beta_i$ separately on Last.FM, BX, ML-1M and AMZ-B, while varying the range of the other parameter. 
The results are presented in Fig.~\ref{fig:parameter}.
It can be observed that: \textbf{(1)} Both $\beta_u$ and $\beta_i$ have an optimal value, and deviating from these values—either higher or lower—leads to a degradation of model performance. \textbf{(2)} The optimal values of $\beta_u$ and $\beta_i$ can differ significantly. For example, on the ML-1M, the optimal $\beta_u$ is $10^{-5}$, while the optimal $\beta_i$ is $10^{-7}$. This indicates that the distributions of different node types may vary considerably, further validating the rationale behind LightKG's node labeling, as demonstrated earlier.

\subsection{Further Exploration of LightKG}
\subsubsection{Semantic Modeling Capabilities} 
Since LightKG represents relations as scalar values, its ability to capture semantic information may be limited. To evaluate this issue, we calculate the variance of aggregation coefficients during GNN propagation - a higher variance suggests better discrimination between different relations. For LightKG, these coefficients are computed via Eq.~\ref{lightkg}, whereas other models derive them from attention weights. As shown in Tab.~\ref{varianc_Comparison}, LightKG consistently achieves the highest variance across all datasets, demonstrating its superior capability to capture global semantic patterns despite its simple scalar-based approach.

\begin{table}[h]
\footnotesize
\setlength{\abovecaptionskip}{3pt}
\centering
\caption{
The variance of aggregation coefficients during GNN.}
\vspace{-0.00in}
\label{varianc_Comparison}
\begin{tabular}{ccccc}
    \toprule
        ~ & Last.FM & ML-1M & BX & AMZ-B  \\ \midrule
        KGAT & 0.0670 & 0.0492 & 0.2248 & 0.0368  \\ 
        KGRec & 0.1273 & 0.0672 & 0.3874 & 0.0514  \\ 
        CL-SDKG & 0.1147 & 0.0694 & 0.3991 & 0.0446  \\ 
        KGIN & 0.1204 & 0.0727 & 0.3433 & 0.0371 \\ \midrule
        LightKG & \textbf{0.5500} & \textbf{0.0967} & \textbf{0.4728} & \textbf{0.0932}  \\ 
        \bottomrule
\end{tabular}

\vspace{-0.4cm}
\end{table}

While LightKG's scalar-based relation encoding offers computational efficiency, it exhibits limited capability to discern fine-grained relational nuances in complex scenarios. For instance, in the ML-1M dataset, the node “Titanic” is linked to 51 other movies via the “film-actor-film” relation. LightKG assigns nearly identical aggregation weights (variance = 0.0009) to these edges, whereas KGAT (variance = 0.0174) demonstrates stronger discriminative power. This limitation is supported by LightKG's suboptimal performance on the MRR metric in the BX dataset. Since MRR relies on the precise ranking position of positive samples, it demands fine-grained discrimination among items with subtle semantic differences—a task inherently more challenging than achieving high Recall.

In summary, LightKG's simple design excels at capturing global semantic information, but this comes at the cost of reduced precision in distinguishing subtle relational details.

\subsubsection{The impact of scalar-based relation} 
In LightKG, scalar values quantitatively measure the relative importance of different relationship types for recommendation tasks. To validate this design, we identify the top-5 most influential relations per dataset based on their scalar weights. The results are shown in Tab.~\ref{scalar}, where the values in parentheses represent the learned relation scalars.

\begin{table}[tp]
\footnotesize
\setlength{\abovecaptionskip}{3pt}
\centering
\caption{The scalar values of top-5 relation from each dataset. For clarity and conciseness, the names of relations have been approximately substituted.}
\vspace{-0.00in}
\label{scalar}
\begin{tabular}{ccccc}
    \toprule
        ~ & Last.FM & ML-1M & BX & AMZ-B  \\ \midrule
        1 & U-I (5.3) & I-U (12.2) & U-I (3.2) & U-I (9.9)  \\
        2 & I-U (5.2) & U-I (7.6) & I-U (2.7) & I-U (9.4)  \\ 
        3 & artist (4.4) & type (6.4) & type (1.2) & char (8.8)  \\ 
        4 & born (3.5) & outfit (6.4) & written (0.9) & written (8.8)  \\ 
        5 & act (2.6) &  act (5.9) & series (0.3) & write (7.7)\\
    \bottomrule
    \end{tabular}
\vspace{-0.5cm}
\end{table}

It can be observed that user-item and item-user interactions ("U-I/I-U") achieve the highest scalar values across all datasets, indicating that direct user-item interactions provide the most influential recommendation signals. 
These results demonstrate that direct user-item interactions consistently outweigh other relational signals in recommender systems. This empirical evidence aligns with and strengthens the arguments presented in \cite{zhang2024does}, which critically examines the actual effectiveness of knowledge graph augmentation in recommender systems.



\section{Related Work}
\subsection{Knowledge-aware Recommender Systems}
Existing KGRSs can be roughly categorized into three types~\cite{jiang2024diffkg}.
    
\smallskip\noindent\textbf{Embedding-based KGRSs}~\cite{zhang1803learning,wang2019multi,li2021kg4vis,zhang2016collaborative,forouzandeh2023new} use the distances and directions in the embedding space to represent relationships between nodes. 
For example, CKE~\cite{zhang2016collaborative} uses TransR to enhance recommendation accuracy by capturing structured semantic relationships between items in the KGs; CFKG~\cite{zhang1803learning} adopts TransE to model user-item interactions, integrating KG to enhance recommendation accuracy and explainability; and KG4Vis~\cite{li2021kg4vis} adapts TransE to model KG relations specifically for automated visualization recommender systems.
    
\smallskip\noindent\textbf{Path-based KGRSs}~\cite{hu2018leveraging,balloccu2023reinforcement,park2022reinforcement,chu2023meta} use methods like random walks to explore item connections in KGs.
For example, RippleNet~\cite{wang2018ripplenet} uncovers latent user-item associations through "ripple propagation".
MCRec~\cite{hu2018leveraging} significantly enhances both recommendation accuracy and diversity by effectively leveraging meta-path contexts, while PGPR~\cite{balloccu2023reinforcement} employs reinforcement learning to optimize the quality of reasoning paths.

\smallskip\noindent\textbf{GNN-based KGRSs} are based on the information aggregation mechanism of GNNs and can be divided into \textbf{supervised methods}~\cite{wang2019KGCN,wang2019knowledge,wang2021learning} and \textbf{self-supervised methods}~\cite{zou2022multi,yang2022knowledge,yang2023knowledge,wu2024efficient,jiang2024diffkg,shi2024self}. 
Regarding supervised methods, KGCN~\cite{wang2019KGCN} and KGNN-LS~\cite{wang2019knowledge} utilize GCN to aggregate neighborhood information of items in KGs. 
Then, KGAT~\cite{wang2019kgat} integrates user-item interactions and KGs, using graph attention to improve recommendations. 
Inspired by the success of self-supervised learning, contrastive learning, as a form of self-supervised learning, has been increasingly integrated into GNN-based methods.
For example, MCCLK~\cite{zou2022multi} aligns knowledge and interaction graphs via cross-view contrastive learning.
KGCL~\cite{yang2022knowledge} reduces data noise through graph contrastive learning~\cite{shi2024self,jiang2024diffkg}.

\subsection{Sparsity Issue in Recommender Systems}
Sparse scenarios arise from limited user-item interaction data, leading to challenges in providing accurate recommendations~\cite{shi2024enhancing}.
Failure to address it will lead to user attrition and significant economic lossess.
To tackle this issue, various data augmentation approaches have been proposed. Auxiliary information, such as KGs~\cite{zou2022multi} and social networks~\cite{wu2020diffnet++}, has been integrated into RSs to enrich the data available for users or items with sparse interactions. Togashi, et al. \cite{togashi2021alleviating} employ an enhanced negative sampling method to mitigate the popularity bias in sparse scenarios. Moreover, incorporating self-supervised learning techniques into RSs has emerged as a promising trend, which addresses the data sparsity issue by extracting additional supervisory signals from raw data~\cite{zou2022multi,yang2022knowledge,yang2023knowledge}.
However, none of these methods have explored what kind of framework is suitable for sparse scenarios.

\subsection{Simplifying GNN for Different Tasks}
Several studies~\cite{gasteiger2018predict,li2019label} have shown that not all components of GNNs are universally beneficial.
For example, SGC~\cite{wu2019simplifying}, a linear simplification of GCNs, achieves superior computational efficiency and parameter economy while maintaining competitive accuracy.
Recent work by Luo et al. \cite{luo2024classic} establishes that traditional GNNs outperform all subsequent architectural modifications on standard node classification benchmarks.
In RSs, LightGCN~\cite{he2020lightgcn}, derived from NGCF~\cite{wang2019neural} through systematic ablation studies, validates that removing nonlinearities and feature transformations can enhance recommendation performance.
These findings collectively underscore a critical design principle: streamlined GNN architectures—when carefully tailored to specific application requirements—consistently deliver enhanced efficiency without compromising effectiveness.

\section{Conclusion and Future Work}
In this work, we empirically reveal the limited or even detrimental effect of complex mechanisms, such as attention mechanism in the existing GNN-based KGRSs.
Motivated by this finding, we propose a simple yet powerful GNN-based KGRS, LightKG, which features a simplified GNN structure with scalar-based relation encoding and linear neighbor-information aggregation mechanism, as well as an efficient contrastive layer to address the over-smoothing issue.  
Extensive experimental results demonstrate LightKG's exceptional performance, achieving superior recommendation accuracy while significantly improving training efficiency.
Our approach offers valuable insights for the design of lightweight and effective GNN-based KGRS.  
For future work, we will further investigate the role of each module of LightKG to increase its interpretability. 
Also, considering that the emergence of large language models~\cite{sun2024large,wang2025re2llm} is generating new types of auxiliary data and also new types of RSs, we plan to adapt LightKG accordingly. 

\section{Acknowledgment}
This research is supported by the State Key Laboratory of Industrial Control Technology, China (Grant No. ICT2024C01), and partially supported by the Fundamental Research Funds for the Central Universities (2025ZFJH02) and the Ministry of Education, Singapore, under its MOE AcRF Tier 1 SUTD Kickstarter Initiative (SKI 2021\_06\_12).

\clearpage
\balance
\bibliographystyle{ACM-Reference-Format}
\bibliography{ref} 

\appendix

\newpage

\section{APPENDIX}
\subsection{The Complete Results of Fig.~\ref{fig:sparse}}
In Section 3.1, we examine the performance of 12 SOTA KGRSs across different sparsity levels on Last.FM, ML-1M and AMZ-B.
\vspace{-0.1cm}
\begin{table}[H]
\footnotesize
\caption{ The Recall@10 of different models across different
sparsity scenarios on Last.FM.}
\vspace{-0.4cm}
  \centering
    \begin{tabular}{ccccc}
    \toprule
    Model & 80\%   & 40\%   & 20\%   & 10\% \\
    \midrule
    CFKG  & 0.2444  & 0.1797  & 0.1082  & 0.0983  \\
    CKE   & 0.2453  & 0.1522  & 0.0544  & 0.0243  \\
    RippleNet & 0.1633  & 0.1192  & 0.0774  & 0.0530  \\
    MCRec & 0.2132  & 0.1243  & 0.0935  & 0.0825  \\
    KGCN  & 0.2149  & 0.1259  & 0.0677  & 0.0425  \\
    KGNNLS & 0.2117  & 0.1281  & 0.0631  & 0.0378  \\
    KGAT  & 0.2583  & 0.1410  & 0.0664  & 0.0202  \\
    KGIN  & 0.2727  & 0.1711  & 0.0770  & 0.0289  \\
    MCCLK & 0.2699  & 0.1759  & 0.0555  & 0.0149  \\
    KGRec & 0.2560  & 0.1758  & 0.0876  & 0.0291  \\
    Diffkg & 0.2520  & 0.1551  & 0.0479  & 0.0177  \\
    CL-SDKG  & 0.2409  & 0.1802  & 0.0621  & 0.0204  \\
    \midrule
    path$_{max}$ & 0.2132  & 0.1243  & 0.0935  & 0.0825  \\
    embedding$_{max}$ & 0.2453  & 0.1797  & \textbf{0.1082 } & \textbf{0.0983 } \\
    GNN$_{max}$ & \textbf{0.2727 } & \textbf{0.1802 } & 0.0876  & 0.0425  \\
    \midrule
    LightKG & 0.2929 & 0.21202 & 0.1012 & 0.0861 \\
    Improve & 7.41\% & 17.66\% & 15.49\% & 102.40\% \\
    \bottomrule
    \end{tabular}%
    \vspace{-0.4cm}
\end{table}%
\begin{table}[H]
\footnotesize
\caption{ The MRR@10 of different models across different
sparsity scenarios on Last.FM.}
\vspace{-0.4cm}
  \centering
    \begin{tabular}{ccccc}
\toprule
Model & 80\%   & 40\%   & 20\%   & 10\% \\
\midrule
CFKG & 0.1100 & 0.0717 & 0.0329 & 0.0304 \\
CKE & 0.1069 & 0.0590 & 0.0200 & 0.0094 \\
RippleNet & 0.0656 & 0.0416 & 0.0259 & 0.0196 \\
MCRec & 0.0941 & 0.0433 & 0.0296 & 0.0231 \\
KGCN & 0.0924 & 0.0474 & 0.0232 & 0.0145 \\
KGNNLS & 0.0891 & 0.0460 & 0.0216 & 0.0110 \\
KGAT & 0.1152 & 0.0529 & 0.0227 & 0.0082 \\
KGIN & 0.1242 & 0.0667 & 0.0294 & 0.0130 \\
MCCLK & 0.1228 & 0.0676 & 0.0204 & 0.0058 \\
KGRec & 0.1118 & 0.0673 & 0.0310 & 0.0125 \\
Diffkg & 0.1192 & 0.0620 & 0.0195 & 0.0059 \\
CL-SDKG & 0.1054 & 0.0571 & 0.0155 & 0.0079 \\
\midrule
path-based$_{max}$ & 0.0941 & 0.0433 & 0.0296 & 0.0231 \\
embedding$_{max}$ & 0.1100 & \textbf{0.0717} & \textbf{0.0329} & \textbf{0.0304} \\
GNN$_{max}$ & \textbf{0.1242} & 0.0676 & 0.0310 & 0.0145 \\
\midrule
LightKG & 0.1350 & 0.0819 & 0.0396 & 0.0267 \\
Improve & 8.61\% & 21.15\% & 27.74\% & 84.14\% \\
\bottomrule
\end{tabular}
\vspace{-0.4cm}
\end{table}%
\begin{table}[H]
\footnotesize
    \caption{The Recall@10 of different models across different
sparsity scenarios on ML-1M.}
  \vspace{-0.4cm}
  \centering
    \begin{tabular}{cccccc}
    \toprule
    Model & 80\%   & 40\%   & 20\%   & 10\%   & 5\% \\
    \midrule
    CFKG  & 0.1862  & 0.1272  & 0.0925  & 0.0615  & 0.0603  \\
    CKE   & 0.1848  & 0.1136  & 0.0760  & 0.0604  & 0.0473  \\
    RippleNet & 0.1590  & 0.1031  & 0.0694  & 0.0619  & 0.0575  \\
    MCRec & 0.1610  & 0.1051  & 0.0690  & 0.0603  & 0.0550  \\
    KGCN  & 0.1594  & 0.1014  & 0.0738  & 0.0604  & 0.0575  \\
    KGNNLS & 0.1592  & 0.0945  & 0.0711  & 0.0601  & 0.0542  \\
    KGAT  & 0.1830  & 0.1124  & 0.0833  & 0.0585  & 0.0311  \\
    KGIN  & 0.1969  & 0.1254  & 0.0802  & 0.0612  & 0.0557  \\
    MCCLK & 0.1853  & 0.1198  & 0.0841  & 0.0523  & 0.0475  \\
    KGRec & 0.1960  & 0.1290  & 0.0934  & 0.0572  & 0.0399  \\
    DiffKG & 0.1846  & 0.0846  & 0.0615  & 0.0512  & 0.0357  \\
    CL-SDKG  & 0.1861  & 0.1122  & 0.0742  & 0.0602  & 0.0445  \\
    \midrule
    path$_{max}$ & 0.1610  & 0.1051  & 0.0694  & \textbf{0.0619 } & 0.0575  \\
    embedding$_{max}$ & 0.1862  & 0.1272  & 0.0925  & 0.0615  & \textbf{0.0603 } \\
    GNN$_{max}$ & \textbf{0.1969 } & \textbf{0.1290 } & \textbf{0.0934 } & 0.0612  & \multicolumn{1}{r}{0.0575 } \\
    \midrule
    LightKG & 0.2015  & 0.1284  & 0.0996  & 0.0699  & 0.0587  \\
    Improve & 2.33\% & -0.47\% & 6.65\% & 14.22\% & 2.16\% \\
    \bottomrule
    \end{tabular}%
\end{table}%
\begin{table}[H]
\footnotesize
    \caption{The MRR@10 of different models across different
sparsity scenarios on ML-1M.}
\vspace{-0.4cm}
  \centering
    \begin{tabular}{cccccc}
\toprule
Model & 80\% & 40\% & 20\% & 10\% & 5\% \\
\midrule
CFKG & 0.3405 & 0.2025 & 0.1487 & 0.1077 & 0.1047 \\
CKE & 0.3457 & 0.1878 & 0.1354 & 0.1092 & 0.0904 \\
RippleNet & 0.3062 & 0.1729 & 0.1250 & 0.1103 & 0.1021 \\
MCRec & 0.3233 & 0.1757 & 0.1245 & 0.1099 & 0.1032 \\
KGCN & 0.3456 & 0.1734 & 0.1307 & 0.1084 & 0.1029 \\
KGNNLS & 0.3051 & 0.1688 & 0.1279 & 0.107 & 0.0989 \\
KGAT & 0.3412 & 0.1870 & 0.1408 & 0.1101 & 0.0647 \\
KGIN & 0.3551 & 0.1976 & 0.1349 & 0.1157 & 0.1004 \\
MCCLK & 0.3474 & 0.1900 & 0.1431 & 0.1063 & 0.0815 \\
KGRec & 0.3570 & 0.2029 & 0.1544 & 0.1086 & 0.0834 \\
DiffKG & 0.3428 & 0.1594 & 0.1202 & 0.0492 & 0.0378 \\
CL-SDKG & 0.3428 & 0.1858 & 0.1329 & 0.1052 & 0.0857 \\
\midrule
path$_{max}$ & 0.3233 & 0.1757 & 0.125 & 0.1103 & 0.1032 \\
embedding$_{max}$ & 0.3457 & 0.2025 & 0.1487 & 0.1077 & \textbf{0.1047} \\
GNN$_{max}$ & \textbf{0.3570} & \textbf{0.2029} & \textbf{0.1544} & \textbf{0.1157} & 0.1029 \\
\midrule
LightKG & 0.3785 & 0.2032 & 0.1491 & 0.1237 & 0.1032 \\
Improve & 6.02\% & 0.14\% & -3.43\% & 5.10\% & 0.09\% \\
\bottomrule
\end{tabular}
\vspace{-0.6cm}
\end{table}%
\begin{table}[H]
\footnotesize
\caption{The Recall@10 of different models across different
sparsity scenarios on AMZ-B.}
\vspace{-0.4cm}
  \centering
\begin{tabular}{ccccc}
\toprule
Model & 80\% & 40\% & 20\% & 10\% \\
\midrule
CFKG & 0.1968 & 0.1357 & 0.1098 & 0.0840 \\
CKE & 0.1979 & 0.1177 & 0.0631 & 0.0237 \\
RippleNet & 0.1561 & 0.1006 & 0.0753 & 0.0446 \\
MCRec & 0.1524 & 0.0983 & 0.0723 & 0.0461 \\
KGCN & 0.1550 & 0.1058 & 0.0757 & 0.0551 \\
KGNNLS & 0.1508 & 0.0948 & 0.0728 & 0.0525 \\
KGAT & 0.1925 & 0.1356 & 0.0679 & 0.0351 \\
KGIN & 0.2090 & 0.1455 & 0.0911 & 0.0765 \\
MCCLK & 0.2025 & 0.1468 & 0.1005 & 0.0671 \\
KGRec & 0.2035 & 0.1448 & 0.0986 & 0.0771 \\
DiffKG & 0.2039 & 0.1355 & 0.0843 & 0.0439 \\
CL-SDKG & 0.2036 & 0.1367 & 0.0947 & 0.0685 \\
\midrule
path$_{max}$ & 0.1561 & 0.1006 & 0.0753 & 0.0461 \\
embedding$_{max}$ & 0.1979 & 0.1357 & \textbf{0.1098} & \textbf{0.0840} \\
GNN$_{max}$ & \textbf{0.2090} & \textbf{0.1468} & 0.1005 & 0.0771 \\
\midrule
LightKG & 0.2120 & 0.1582 & 0.1148 & 0.0931 \\
Improve & 1.44\% & 7.77\% & 14.23\% & 20.75\% \\
\bottomrule
\end{tabular}
\vspace{-0.6cm}
\end{table}%
\begin{table}[H]
\footnotesize
    \caption{The MRR@10 of different models across different
sparsity scenarios on AMZ-B.}
\vspace{-0.4cm}
  \centering
\begin{tabular}{ccccc}
\toprule
Model & 80\% & 40\% & 20\% & 10\% \\
\midrule
CFKG & 0.0987 & 0.0632 & 0.0515 & 0.0422 \\
CKE & 0.1037 & 0.0575 & 0.0266 & 0.0146 \\
RippleNet & 0.0838 & 0.0484 & 0.0422 & 0.0257 \\
MCRec & 0.0791 & 0.0472 & 0.0415 & 0.0274 \\
KGCN & 0.0738 & 0.0501 & 0.0418 & 0.0309 \\
KGNNLS & 0.0750 & 0.0482 & 0.0396 & 0.0247 \\
KGAT & 0.0997 & 0.0652 & 0.0283 & 0.0169 \\
KGIN & 0.1099 & 0.0722 & 0.0452 & 0.0407 \\
MCCLK & 0.1065 & 0.0715 & 0.5050 & 0.0389 \\
KGRec & 0.1094 & 0.0707 & 0.0483 & 0.0415 \\
DiffKG & 0.1116 & 0.0622 & 0.0457 & 0.0286 \\
CL-SDKG & 0.1134 & 0.0635 & 0.0473 & 0.0404 \\
\midrule
path$_{max}$ & 0.0838 & 0.0484 & 0.0422 & 0.0274 \\
embedding$_{max}$ & 0.1037 & 0.0632 & \textbf{0.0515} & \textbf{0.042}2 \\
GNN$_{max}$ & \textbf{0.1134} & \textbf{0.0722} & 0.0505 & 0.0415 \\
\midrule
LightKG & 0.1173 & 0.0762 & 0.0577 & 0.0472 \\
Improve & 3.44\% & 5.54\% & 14.26\% & 13.73\% \\
\bottomrule
\end{tabular}
\vspace{-0.3cm}
\end{table}%
Due to the sparsity of the BX dataset, it is difficult to further reduce the sampling rate, making it infeasible for us to conduct experiments on it.  

\subsection{The Complete Results of Tab.~\ref{complexivity}}
We show the complete results of Tab.~\ref{complexivity}. The sampling ratio is set 10\% for Last.FM and AMZ-B, 5\% for ML-1M. Since the BX dataset is already extremely sparse, 80\% sampling ratio is sufficient. \textbf{The Pearson correlation coefficients on four datasets are all negative}.
\vspace{-0.5cm}
\begin{table}[H]
\setlength{\abovecaptionskip}{3pt}
\centering
\footnotesize
  \caption{The results of complexity analysis.}
\vspace{-1pt}
\resizebox{0.975\linewidth}{!}{
    \begin{tabular}{cccccc}
    \toprule
    Model & Time Complexity (ranked from low to high) & Last.FM & ML-1M & Amz-B & BX \\
    \midrule
    KGCN  & $\mathcal{O}(d(|\mathcal{G}_{KG}|+|\mathcal{R}|\times|\mathcal{U}|+|\mathcal{I}|))$ & 0.0425 & 0.0575 & 0.0551 & 0.0867 \\
    KGNNLS & $\mathcal{O}(d(|\mathcal{G}_{KG}|+|\mathcal{V}|+|\mathcal{U}|\times|\mathcal{R}|))$ & 0.0378 & 0.0542 & 0.0525 & 0.0731 \\
    KGIN  & $\mathcal{O}(d|\mathcal{G}_{KG}| + d^2|\mathcal{G}_{UI}|)$ & 0.0289 & 0.0557 & 0.0765 & 0.0801 \\
    KGRec & $\mathcal{O}(d^2|\mathcal{G}_{KG}| + d|\mathcal{G}_{UI}|)$ & 0.0291 & 0.0398 & 0.0771 & 0.1033 \\
    DiffKG & $\mathcal{O}(d^2|\mathcal{G}_{KG}| + d|\mathcal{G}_{UI}|)$ & 0.0177 & 0.0357 & 0.0439 & 0.0581 \\
    CL-SDKG  & $\mathcal{O}(d^2(|\mathcal{G}_{KG}|+|\mathcal{U}|) + d|\mathcal{G}_{UI}|)$ & 0.0204 & 0.0445 & 0.0685 & 0.0924 \\
    KGAT  & $\mathcal{O}(d^2(|\mathcal{G}_{KG}|+|\mathcal{G}_{UI}|+|\mathcal{U}|+|\mathcal{I}|+|\mathcal{V}|))$ & 0.0202 & 0.0311 & 0.0351 & 0.0499 \\
    MCCLK & $\mathcal{O}(d^3|\mathcal{G}_{KG}|+d|\mathcal{G}_{UI}|)$ & 0.0149 & 0.0475 & 0.0671 & 0.0607 \\
    \midrule
    Correlation &       & -0.9374 & -0.6682 & -0.1145 & -0.4836 \\
    \bottomrule
    \end{tabular}%
    }
\vspace{-0.3cm}
\end{table}%

\subsection{The Complete Results of Tab.~\ref{tab:attention_moved}}
Last, we present the rest of experimental results with the attention mechanism removed. Noticing that removing the attention mechanism leads to a slight decline in performance on the AMZ-B, we attribute this to the method we used to remove the attention mechanism being too crude, which damaged the model's structure.
This slight decline does not affect our conclusion: the attention mechanism is useless even harmful in GNN-based KGRSs.
\vspace{-0.2cm}
\begin{table}[H]
\footnotesize
\caption{The MRR@10 after removing the attention mechanism on Last.FM. }
\vspace{-0.4cm}
  \centering
    \begin{tabular}{ccccc}
    \toprule
model            & 80\%             & 40\%             & 20\%            & 10\%             \\ \midrule
KGAT             & 0.1152          & \textbf{0.0529}           & 0.0227          & 0.0082          \\
KGAT$_{a-}$     & \textbf{0.1172}          & 0.0475           & \textbf{0.0264}          & \textbf{0.0096}          \\ \midrule
KGIN             & \textbf{0.1243}          & \textbf{0.0667}           & 0.0294          & 0.0130           \\
KGIN$_{a-}$     & 0.1241          & 0.0654           & \textbf{0.0301}          & \textbf{0.0138}          \\ \midrule
MCCLK            & \textbf{0.1228}          & \textbf{0.0676}           & 0.0204          & 0.0058          \\
MCCLK$_{a-}$    & 0.1215          & 0.0668           & \textbf{0.0209}          & \textbf{0.0059}          \\ \midrule
KGRec            & 0.1117          & \textbf{0.0673}           & 0.0310           & 0.0125          \\
KGRec$_{a-}$    & \textbf{0.1140}          & 0.0669           & \textbf{0.0311}          & 0.0125          \\ \midrule
DiffKG           & \textbf{0.1192}          & 0.0620            & 0.0195          & 0.0059          \\
DiffKG$_{a-}$   & 0.1177          & \textbf{0.0627}           & 0.0195          & \textbf{0.0067}          \\ \midrule
CL-SDKG             & 0.1054          & \textbf{0.0571}           & 0.0155          & 0.0079          \\
CL-SDKG$_{a-}$    & \textbf{0.1059}          & 0.0562           & \textbf{0.0157}          & 0.0079          \\ \midrule
Average          & 0.1164          & 0.0623           & 0.0231          & 0.0089          \\
Average$_{a-}$  & 0.1167          & 0.0609           & 0.0240          & 0.0094          \\
Improve & 0.26\% & -2.17\% & 3.75\% & 5.82\% \\
\bottomrule
\end{tabular}
\vspace{-0.4cm}
\end{table}%
\begin{table}[H]
\footnotesize
\caption{The Recall@10 after removing the attention mechanism on ML-1M. }
\vspace{-0.4cm}
  \centering
    \begin{tabular}{cccccc}
    \toprule
        Model & 80\% & 40\% & 20\% & 10\% & 5\% \\
    \midrule
    KGAT  & 0.1830  & 0.1124  & \textbf{0.0833}  & 0.0585  & 0.0311  \\
    KGAT$_{a-}$ & \textbf{0.1837}  & \textbf{0.1152}  & 0.0795  & \textbf{0.0611}  & \textbf{0.0317}  \\
    \midrule
    KGIN  & 0.1969  & \textbf{0.1254}  & \textbf{0.0802}  & \textbf{0.0612}  & \textbf{0.0557}  \\
    KGIN$_{a-}$ & \textbf{0.1971}  & 0.1250  & 0.0799  & 0.0607  & 0.0554  \\
    \midrule
    MCCLK & 0.1853  & \textbf{0.1198}  & 0.0841  & \textbf{0.0723}  & 0.0475  \\
    MCCLK$_{a-}$ & \textbf{0.1860}  & 0.1193  & 0.0841  & 0.0693  & \textbf{0.0478}  \\
    \midrule
    KGRec & 0.1960  & 0.1290  & 0.0934  & 0.0572  & 0.0399  \\
    KGRec$_{a-}$ & \textbf{0.1966}  & \textbf{0.1293}  & \textbf{0.0942}  & \textbf{0.0600}  & 0.0399  \\
    \midrule
    DiffKG & 0.1846  & 0.0846  & \textbf{0.0615}  & 0.0512  & 0.0357  \\
    DiffKG$_{a-}$ & \textbf{0.1892}  & \textbf{0.0877}  & 0.0611  & \textbf{0.0557}  & \textbf{0.0391}  \\
    \midrule
    CL-SDKG  & 0.1861  & 0.1122  & 0.0742  & 0.0602  & 0.0445  \\
    CL-SDKG$_{a-}$ & \textbf{0.1881}  & \textbf{0.1134}  & \textbf{0.0795}  & \textbf{0.0611}  & \textbf{0.0473}  \\
    \midrule
    Average & 0.1887  & 0.1139  & 0.0795  & 0.0601  & 0.0424  \\
    Average$_{a-}$ & 0.1901  & 0.1150  & 0.0797  & 0.0613  & 0.0435  \\
    Improve & 0.76\% & 0.95\% & 0.33\% & 2.02\% & 2.70\% \\
    \bottomrule
    \end{tabular}%
\end{table}%
\begin{table}[H]
\footnotesize
\caption{The MRR@10 after removing the attention mechanism on ML-1M. }
\vspace{-0.4cm}
  \centering
    \begin{tabular}{cccccc}
\toprule
model            & 80\%             & 40\%             & 20\%             & 10\%             & 5\%            \\
\midrule
KGAT             & 0.3412          & 0.187           & \textbf{0.1408}          & 0.1101          & \textbf{0.0647}          \\
KGAT$_{a-}$           & \textbf{0.3413}          & \textbf{0.1885}          & 0.1349          & 0.1101          & 0.0638          \\ \midrule
KGIN             & 0.3551          & \textbf{0.1976}          & 0.1349          & \textbf{0.1157}          & \textbf{0.1004}          \\
KGIN$_{a-}$          & \textbf{0.3596}          & 0.1950          & \textbf{0.1362}          & 0.1152          & 0.0996          \\ \midrule
MCCLK            & 0.3474          & \textbf{0.1900}            & 0.1431          & 0.1063          & 0.0815          \\
MCCLK$_{a-}$         & \textbf{0.3475}          & 0.1898          & \textbf{0.1434}          & \textbf{0.1081}          & \textbf{0.0823}          \\ \midrule
KGRec            & \textbf{0.3570}           & 0.2029          & 0.1544          & 0.1086          & 0.0834          \\
KGRec$_{a-}$         & 0.3567          & \textbf{0.2070}          & \textbf{0.1565}          & \textbf{0.1124}          & 0.0834          \\ \midrule
DiffKG           & 0.3428          & 0.1594          & 0.1202          & 0.0492          & 0.0378          \\
DiffKG$_{a-}$        & \textbf{0.3441}          & \textbf{0.1616}          & \textbf{0.1215}          & \textbf{0.0509}          & \textbf{0.0385}          \\ \midrule
CL-SDKG             & 0.3428          & 0.1858          & 0.1329          & 0.1052          & 0.0857          \\
CL-SDKG$_{a-}$          & \textbf{0.3444}          & \textbf{0.1872}          & \textbf{0.1338}          & \textbf{0.1067}          & \textbf{0.0877}          \\ \midrule
Average          & 0.3477          & 0.1871          & 0.1377          & 0.0992          & 0.0756          \\
Average$_{a-}$  & 0.3489          & 0.1882          & 0.1377          & 0.1006          & 0.0759          \\
Improve & 0.35\% & 0.57\% & 0.00\% & 1.39\% & 0.40\% \\
\bottomrule
\end{tabular}
\vspace{-0.5cm}
\end{table}%
\begin{table}[H]
\footnotesize
\caption{The Recall@10 after removing the attention mechanism on AMZ-B. }
\vspace{-0.4cm}
  \centering
    \begin{tabular}{ccccc}
    \toprule
Model           & 80\% & 40\% & 20\% & 10\%     \\
\midrule
KGAT            & \textbf{0.1925} & \textbf{0.1356} & \textbf{0.0679} & \textbf{0.0351}  \\
KGAT$_{a-}$    & 0.1872 & 0.1342 & 0.0666 & 0.0316  \\
\midrule
KGIN            & \textbf{0.2090} & 0.1455 & \textbf{0.0911} & \textbf{0.0765}  \\
KGIN$_{a-}$    & 0.2081 & \textbf{0.1466} & 0.0901 & 0.0738  \\
\midrule
MCCLK           & \textbf{0.2025} & 0.1468 & 0.1005 & 0.0671  \\
MCCLK$_{a-}$   & 0.2016 & \textbf{0.1477} & \textbf{0.1042} & 0.0671  \\
\midrule
KGRec           & 0.2035 & 0.1448 & \textbf{0.0986} & 0.0771  \\
KGRec$_{a-}$   & \textbf{0.2041} & \textbf{0.1476} & 0.0976 & \textbf{0.0773}  \\
\midrule
DiffKG          & 0.2039 & 0.1355 & 0.0843 & 0.0439  \\
DiffKG$_{a-}$  & \textbf{0.2098} & \textbf{0.1394} & \textbf{0.0875} & \textbf{0.0477}  \\
\midrule
CL-SDKG            & 0.2036 & 0.1367 & 0.0947 & 0.0685  \\
CL-SDKG$_{a-}$    & \textbf{0.2050} & \textbf{0.1379} & \textbf{0.0964} & \textbf{0.0697}  \\
\midrule
Average         & 0.2025 & 0.1408 & 0.0895 & 0.0614  \\
Average$_{a-}$ & 0.2026 & 0.1422 & 0.0904 & 0.0612  \\
Improve         & 0.07\% & 1.01\% & 0.99\% & -0.28\% \\
\bottomrule
\end{tabular}
\vspace{-0.5cm}
\end{table}%
\begin{table}[H]
\footnotesize
\caption{The MRR@10 after removing the attention mechanism on AMZ-B. }
\vspace{-0.4cm}
  \centering
    \begin{tabular}{ccccc}
    \toprule
model            & 80\%              & 40\%              & 20\%             & 10\%              \\ \midrule
KGAT             & \textbf{0.0997}           & \textbf{0.0652}           & \textbf{0.0283}          & \textbf{0.0169}           \\
KGAT$_{a-}$     & 0.0956           & 0.0644           & 0.0275          & 0.0143           \\ \midrule
KGIN             & \textbf{0.1099}           & 0.0722           & 0.0452          & \textbf{0.0407}           \\
KGIN$_{a-}$     & 0.1097           & \textbf{0.0731}           & 0.0452          & 0.0391           \\ \midrule
MCCLK            & 0.1065           & \textbf{0.0715}           & 0.0505          & 0.0389           \\
MCCLK$_{a-}$    & \textbf{0.1067 }          & 0.0711           & \textbf{0.0508}          & 0.0389           \\ \midrule
KGRec            & \textbf{0.1094}           & \textbf{0.0707}           & \textbf{0.0483}          & \textbf{0.0415}           \\
KGRec$_{a-}$    & 0.1083           & 0.0699           & 0.0477          & 0.0406           \\ \midrule
DiffKG           & 0.1116           & 0.0622           & 0.0457          & 0.0286           \\
DiffKG$_{a-}$   & \textbf{0.1159}           & \textbf{0.0627}           & \textbf{0.0462}          & \textbf{0.0291}           \\ \midrule
CL-SDKG             & 0.1134           & \textbf{0.0635}           & 0.0473          & 0.0404           \\
CL-SDKG$_{a-}$     & \textbf{0.1138}           & 0.0631           & \textbf{0.0479}          & \textbf{0.0408}           \\ \midrule
Average          & 0.1084           & 0.0676           & 0.0442          & 0.0345           \\
Average$_{a-}$  & 0.1083           & 0.0674           & 0.0442          & 0.0338           \\
Improve & -0.08\% & -0.25\% & 0.00\% & -2.03\% \\
\bottomrule
\end{tabular}
\end{table}%

\clearpage


\end{document}